\documentclass{aa}

\usepackage{graphicx}
\usepackage{txfonts}
\usepackage{url}
\usepackage{natbib}
\usepackage{color}
\bibpunct{(}{)}{;}{a}{}{,}
\begin{document}

\title{The optical properties of dust: the effects of composition, size, and structure}

\author{N. Ysard\inst{\ref{inst1}}
\and A.P. Jones\inst{\ref{inst1}}
\and K. Demyk\inst{\ref{inst2}}
\and T. Bout\'eraon\inst{\ref{inst1}}
\and M. Koehler\inst{\ref{inst1}}}

\institute{Institut d'Astrophysique Spatiale, CNRS, Univ. Paris-Sud, Universit{\'e} Paris-Saclay, B{\^a}t. 121, 91405 Orsay cedex, France \label{inst1} 
\and Institut de Recherche en Astrophysique et Plan{\'e}tologie, CNRS, Universit{\'e} de Toulouse, 9 avenue du Colonel Roche, 31028 Toulouse Cedex 4, France\label{inst2}
\\ \email{nathalie.ysard@ias.u-psud.fr}}

\abstract
{Dust grains are determinant for setting the chemical, physical, dynamical, and radiative properties of all the media in which they are present. Their influence depends on the grain composition, size, and geometrical structure which vary throughout the lifecycle of dust. Particularly grain growth arises in dense molecular clouds and protoplanetary disks as traced by an enhancement of the dust far-IR emissivity and by the effects of cloudshine and coreshine.}
{Our aim is to investigate the imprint of the grain characteristics on the dust unpolarised optical properties from the visible to the far-IR wavelengths for isolated grains as well as for aggregates.}
{Using optical constants for both carbonaceous and silicate materials, we derive the absorption and scattering efficiencies, the asymmetry factor of the phase function, the single scattering albedo, and the mass opacity for isolated grains and aggregates, using either the Mie theory or the Discrete Dipole Approximation (DDA). We investigate the effects of the size, porosity and shape of the grains, and of the monomers constituting the aggregates, on the optical properties. Besides, for aggregates we study the influence of the number of monomers and of mixing monomer sizes. }
{Grain structure changes result in optical property variations at all wavelengths. Porosity, grain elongation, as well as aggregation all produce an increase in the far-IR opacity. The spectral dependence of this increase depends on the nature of the material composing the grain: it is independent of the wavelength for insulators but not for conductors. In the case of aggregates, the far-IR increase does not depend on the monomer size and saturates for aggregates containing 6 or more monomers. In the visible and near-IR, the aggregate behaviour is reminiscent of a compact sphere of the same mass whereas at longer wavelengths, it is closer to the effect of porosity. Finally, for silicates, the mid-IR spectral feature at 18 $\mu$m is more sensitive to the details of the grain structure than the 10 $\mu$m feature.}
{Dust optical properties, from the visible to the far-IR, are highly dependent upon the grain composition, size, and structure. This study provides a basis for understanding the range of variations achievable as a result of varying the grain characteristics. It emphasises the importance of considering the detailed grain structure in determining the dust optical properties and of using exact methods because approximate methods are unable to reproduce the entire range of the observed variations at all wavelengths.}

   \authorrunning{Ysard et al.}
\titlerunning{}
\maketitle
%

\section{Introduction}
\label{introduction}

Dust grains are ubiquitous in all astrophysical environments, from the Solar System and protoplanetary disks to interstellar and intergalatic clouds, and their influence on the radiative properties of all these very diverse media is always significant through the absorption, scattering, and (non-)thermal re-emission of starlight. They are also a major player in the determination of the interstellar gas temperature through photoelectric emission or gas-grain collisions \citep[e.g.][]{Hollenbach1989, Weingartner2001}. Similarly grains have a great influence on the chemical complexity in the interstellar medium (ISM): indeed, the r\^ole of grain-surface reactions is crucial to understand the formation of some very common molecules, such as H$_2$, and of more complex molecules \citep[see for instance][]{Williams2005, Wakelam2017}. The grain radiative properties and their catalytic efficiency are, at least, reliant on the grain size distribution and chemical composition. A long-standing issue is their detailed geometrical structure: are they isolated particles or aggregates? Are the monomers spherical or spheroidal? Are the monomers porous or compact? Is their surface smooth or rough? These structural properties are known to greatly influence the grain optical properties as already demonstrated by many studies of interstellar dust \cite[e.g.][]{Bazell1990, Kozasa1992, Stognienko1995, Fogel1998, Voshchinnikov2000, Shen2008, Koehler2011, Ormel2011, Koehler2012, Kataoka2014, Koehler2015, Min2016}, cometary dust \citep[e.g.][]{Okamoto1998, Kimura2003, Kimura2016}, or terrestrial aerosols of interest in the study of climate change \citep[e.g.][]{Kemppinen2015, Liu2015, Wu2016, Doner2017}. The intent of the present paper is to systematically, if not exhaustively, investigate the imprint of grain structure on the dust unpolarised optical properties from the visible to the far-IR wavelengths for isolated grains as well as aggregates. Variations observed in the optical properties with grain structure are also known to depend on whether the considered material is highly absorbing or not \citep[e.g.][]{BHMIE, Voshchinnikov2000}. Our study will thus use both insulating and conducting materials.

The aim of this study is not to provide an advanced grain model directly applicable to the dense interstellar medium or protoplanetary disks but to understand in detail which parameters are capable of modifying the dust optical properties and are therefore at the origin of the variations in the astronomical dust observables. These variations include a decrease in grain temperature, an increase in the far-IR emissivity, and variations in the spectral index which are usually explained by increasing grain size and/or fluffiness/surface irregularity/coagulation/ice coating \citep[e.g.][]{Ossenkopf1993, Stepnik2003, Ormel2011, Ridderstad2010, Ysard2013, Koehler2015, Min2016}. These variations are often accompanied by an increase in the scattering efficiency from the visible to the mid-IR known as cloudshine and coreshine \citep[e.g.][]{Mattila1970, Lehtinen1996, Foster2006, Pagani2010, Paladini2014}.

This paper is organised as follows. Section~\ref{grain_description} describes the grain compositions, sizes, and structures considered and how their optical properties are calculated. Section~\ref{extinction_efficiency_mass_opacity} explores the effect of changing all these parameters on the extinction efficiency and dust mass opacity. In Sect.~\ref{silicate_midIR_feature}, changes in the silicate mid-IR features are discussed. Albedo variations are discussed in Sect.~\ref{scattering_properties}. Section~\ref{DDA_vs_EMT} presents the comparison between exact calculations of the optical properties and approximate methods. Concluding remarks are given in Sect.~\ref{conclusion}.

\section{Grain description}
\label{grain_description}

The dust optical properties (absorption and scattering efficiencies/cross-sections) and mass opacity (extinction cross-section per unit mass) depend on three parameters: (i) the material complex refractive index $m = n + ik$ (optical constants); (ii) the grain size; and (iii) the grain structure (core/mantle, multi-layer, aggregate...). In this section, we describe all the variations in these parameters which are considered in this study and the methods used to translate them into optical properties (absorption and scattering efficiencies).

\subsection{Optical constants}
\label{optical_constants}

Our starting point is the global dust modelling framework THEMIS\footnote{See \url{http://www.ias.u-psud.fr/themis/}.} (The Heterogeneous dust Evolution Model for Interstellar Solids), which is briefly summarised in \citet{Jones2017}\footnote{For the full details of the model see: \citet{Jones2012a, Jones2012b, Jones2012c, Jones2013, Koehler2014, Koehler2015, Ysard2015, Jones2016, Ysard2016}.}. Two main sets of optical constants are included in THEMIS: amorphous magnesium-rich silicates with metallic iron and iron sulphide nano-inclusions and amorphous hydrocarbon grains.

\begin{figure}[!t]
\centerline{\includegraphics[width=0.52\textwidth]{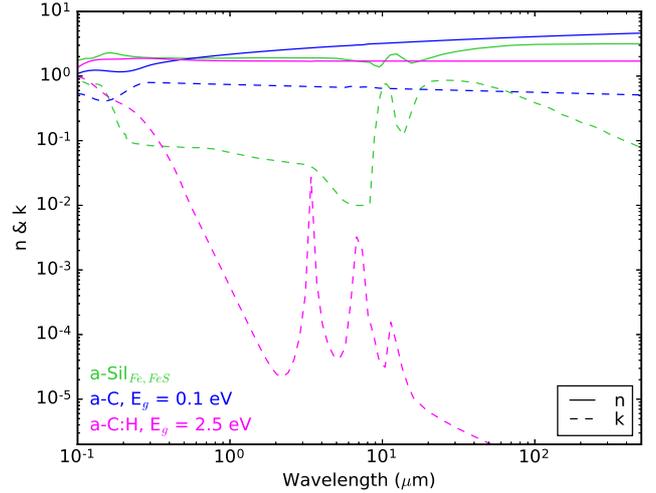}}
\caption{Optical constants of silicates (green), a-C with $E_g = 0.1$~eV (blue), and a-C:H with $E_g = 2.5$~eV (fuchsia). The real, $n$, and complex, $k$, parts of the complex refractive index are showed by the solid and dashed lines, respectively.}
\label{fig_optical_constants} 
\end{figure}

As described in \citet{Koehler2014}, the silicates are assumed to be a 50-50\% mixture of grains with the normative compositions of forsterite and enstatite, the main difference between these two materials being the 10 and 20~$\mu$m band profiles. As the differences are small, we only consider forsterite-type amorphous silicates in this study, which have a density of 2.95~g/cm$^3$. In the following, grains built with these optical constants will be referred to as a-Sil$_{{\rm Fe,FeS}}$ grains (see green lines in Fig.~\ref{fig_optical_constants}).

The THEMIS hydrocarbon dust component, a-C(:H), is described in \citet{Jones2012a, Jones2012b, Jones2012c}. The optical constants of these semi-conducting materials are built using the extended Random Covalent Network (eRCN) and the Defective Graphite (DG) models. These two models allow us to derive the complex refractive indices of a-C(:H) as a function of their Tauc band gap, $E_g$, and the particle size, $a$\footnote{In this study, we only consider grains with sizes $\geqslant 0.1~\mu$m for which size effects on optical constants are negligible.}. In order to show the biggest possible variations, we use the optical properties of two extreme cases in this study: (i) H-poor/aromatic-rich a-C grains with $E_g = 0.1$~eV (blue lines in Fig.~\ref{fig_optical_constants}) and (ii) H-rich/aliphatic rich a-C:H grains with $E_g = 2.5$~eV (fuchsia lines in Fig.~\ref{fig_optical_constants}). According to \citet{Jones2012c}, the a-C(:H) material density can be related to the band gap through the following relation: $\rho(E_g) \approx 1.3 + 0.4 {\rm exp}[-(E_g + 0.2)]$~g/cm$^3$, leading to $\rho = 1.60$ and 1.33~g/cm$^3$ for the a-C and a-C:H materials, respectively. Of course, the growth of pure big a-C(:H) carbon grains in the ISM seems highly improbable as we would rather expect mixing with the silicate dust component but our primary aim here is to investigate whether the variations in the optical properties due to grain shape are subordinate to the material nature or not.

\subsection{Grain sizes and structures}
\label{sizes_and_structures}

The size and structure of interstellar grains are still a matter of debate for the various ISM phases. In order to be as exhaustive as possible, we consider as many variations in these two parameters as possible. Our calculations can be split in two main grain types: isolated grains and aggregates.

For the isolated grains, we consider three types of variations:
\begin{itemize}
\item compact spherical grains with grain radii $a = 0.05$ to 5~$\mu$m;
\item porous spherical grains with $a = 0.1~\mu$m and a porosity degree $P = 10$, 20, 30, 40, and 50\% where a porosity degree of $x\%$ corresponds to the random removal of $x\%$ of the material composing a compact sphere of the same radius\footnote{Following the grain structure definition required by the \texttt{ddscat} routine \citep[][, see Sect.~\ref{optical_property_calculations} for details]{DDA}, porosity is modelled by removing one dipole at a time, not an adjacent group of dipoles.};
\item compact oblate and prolate grains with aspect ratios of 2: the principal axis of the grains have a length of $2a$ and the two other axis of $2b$ with $b/a = 2$ for oblates and $b/a = 1/2$ for prolates, corresponding to volume equivalent radii $a_V = 0.05, 0.1$ and 1~$\mu$m, where $a_V = (b^2a)^{1/3}$.
\end{itemize}

\begin{figure}[!t]
\centerline{\begin{tabular}{cc}
\includegraphics[width=0.23\textwidth]{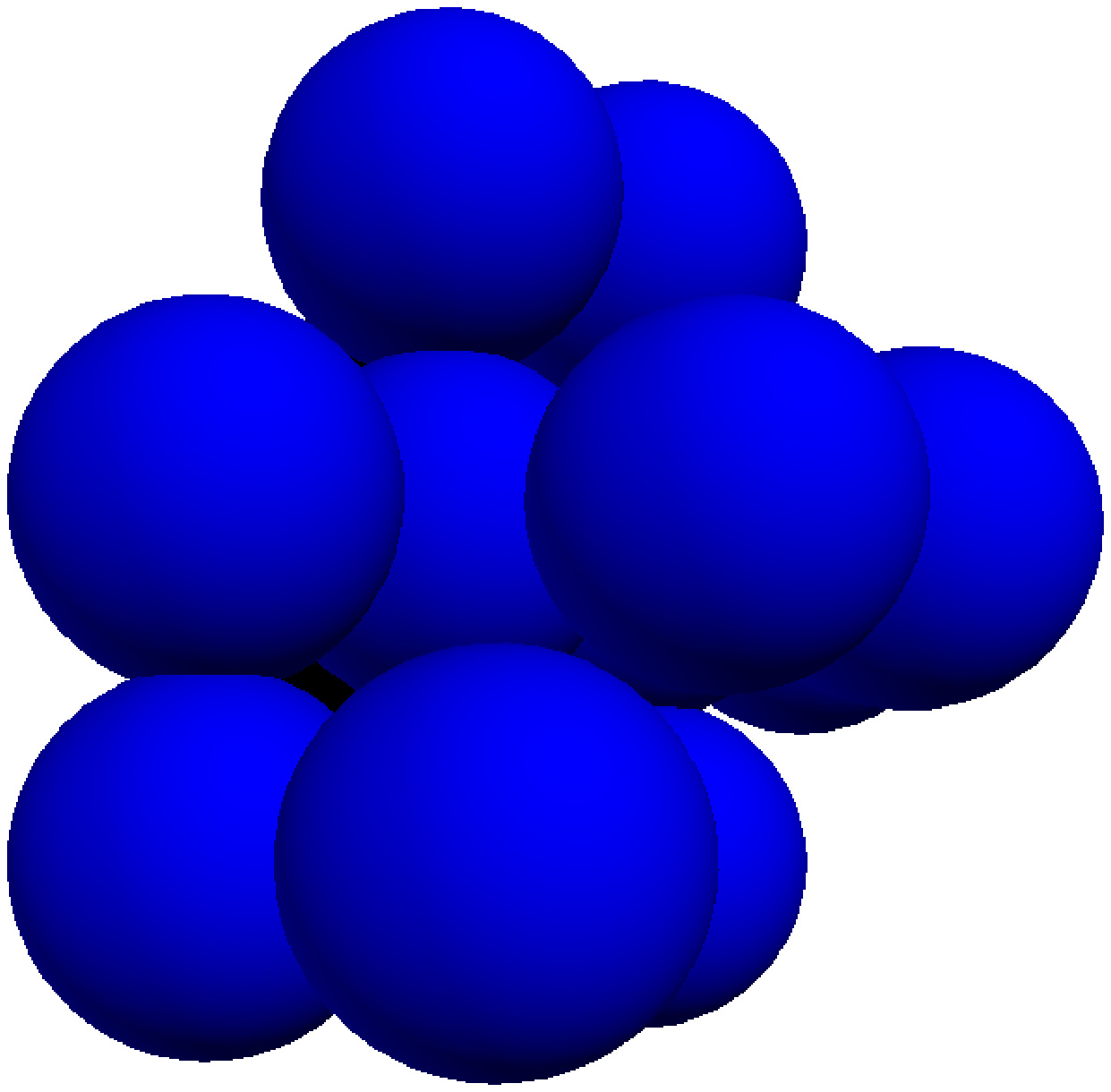} & \includegraphics[width=0.23\textwidth]{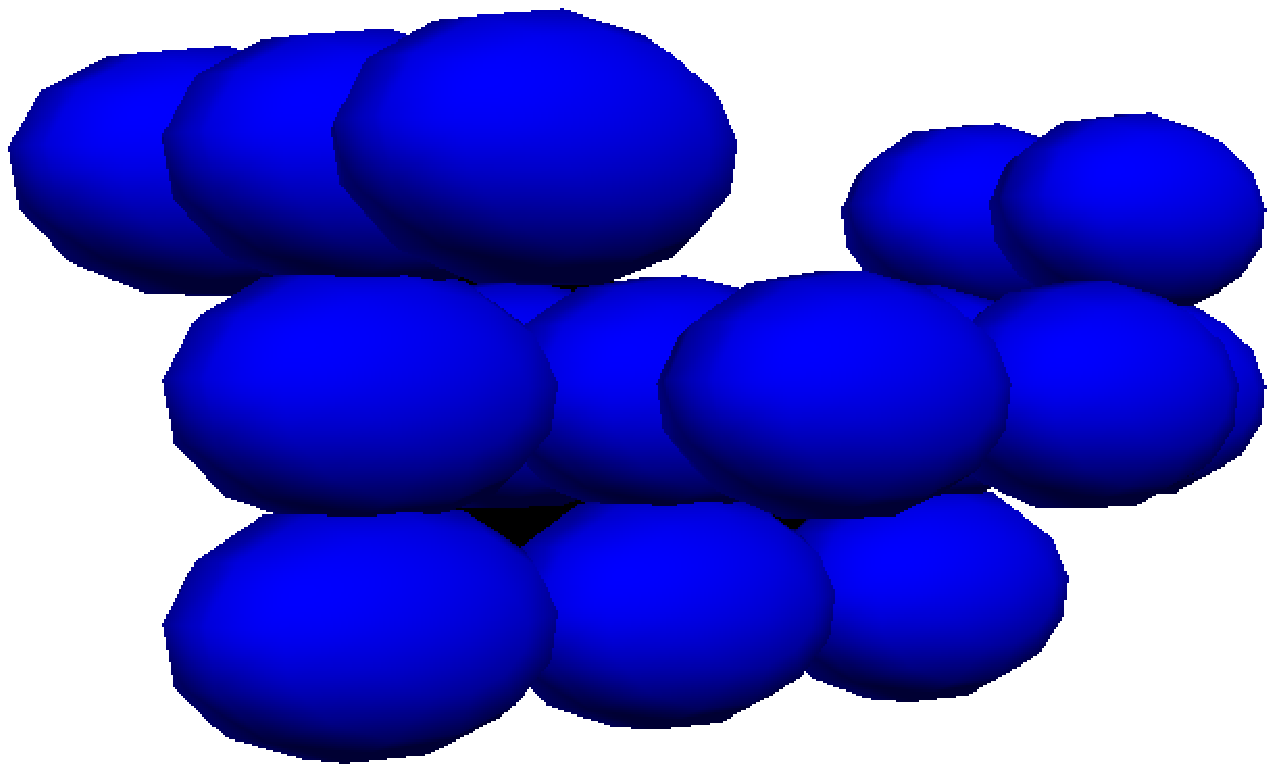}
\end{tabular}}
\caption{Two examples of aggregate shapes (arbitrary scales). Left: aggregate made of $N = 12$ compact spherical monomers with $a_0 = 0.1~\mu$m. Right: aggregate made of $N = 16$ compact prolate monomers with $a_0 = 0.05~\mu$m and aspect ratios of 2.}
\label{fig_structures}
\end{figure}

For the aggregates, we follow the approach of \citet{Koehler2011}. These authors considered aggregates made of $N$ compact spherical monomers of constant radius $a_0 = 0.1~\mu$m located on a cubic grid, composed of amorphous silicates or carbons. They studied the influence of the number $N$ of monomers composing the aggregates and of the contact area between the monomers on the grain near to far-IR extinction efficiency. They found an increase in the efficiency when increasing the contact area up to a diameter of about the monomer radius. Increasing further the contact area had no further effect. Similar results were also found by \citet{Xing1997} for amorphous silicates and glassy carbons and by \citet{Yon2015} and \citet{Doner2017bis} for soot aggregates in the near-IR. In the following, in order to maximise the variations in the optical properties, we always assume the contact area of maximum effect for all the considered aggregates with $N = [2, 4, 6, 8, 10, 12, 14, 16]$:
\begin{itemize}
\item aggregates made of $N$ compact spherical monomers of radius $a_0 = 0.05, 0.1, 0.5$, and 1~$\mu$m;
\item aggregates made of $N$ porous spherical monomers of radius $a_0 = 0.1~\mu$m with $P = 20\%$;
\item aggregates made of $N$ compact spheroid monomers of mass/volume equivalent radius $a_V = 0.05~\mu$m: we consider oblate and prolate monomers, both with aspect ratios of 2, and for the sake of simplicity, all the spheroids have their principal axes aligned along the same axis;
\item aggregates made of four compact spherical monomers with two different radii: $a_1 = 0.05~\mu$m and $a_2 = 0.1$ or 0.5~$\mu$m. 
\end{itemize}
Aggregates with more than two monomers can have several structures. In this study, following \citet{Koehler2011}, we average our results over 10 randomly chosen aggregate shapes for $N \geqslant 6$ and exclude the most compact and most elongated shapes when $N = 4$. One way of characterising the shape is through the grain fractal dimension $D_f$, which can be defined as \citep{Bazell1990, Jones2011}:
\begin{equation}
N(r) \; {\rm or} \; M(r) \propto r^{D_f}, \; 1 \leqslant D_f \leqslant 3,
\end{equation}
where $N(r)$ and $M(r)$ are the number and mass, respectively, of particles within a given radius $r$ measured from a reference point (the mass centre of the aggregate in our case). If $D_f = 1$, the aggregate is linear and if $D_f = 3$, it is compact. As explained in \citet{Kozasa1992} two extreme cases of aggregates can be considered: BPCA (Ballistic Particle Cluster Aggregates) with $D_f = 3$ and BCCA (Ballistic Cluster Cluster Aggregates) with $D_f = 2$. In their simulations, the aggregates were produced by ``shooting projectiles onto a target randomly one at a time'': in the BPCA case, the projectile was a constituent particle, whereas in the BCCA case, the projectile was an aggregate of the same mass as the target but of different shape. In order to reduce the computation time, we only consider an intermediate case in our study and build the aggregates so that they obey $D_f \sim 2.5$. Aggregate shape examples are shown in Fig.~\ref{fig_structures}.

\subsection{Optical property calculations}
\label{optical_property_calculations}

Two methods are used to derive the optical properties from the optical constants. For compact spherical grains, we use the Fortran 90 version of the Mie theory routine \texttt{BHMIE} described in \citet{BHMIE}. For more complex grains, we use the Discrete Dipole Approximation \citep[DDA,][]{Purcell1973, Draine1988} utilising the publicly available \texttt{ddscat} routine\footnote{\url{https://www.astro.princeton.edu/~draine/DDSCAT.7.3.html}} described in \citet{DDA}. In DDA, the grain is assumed to be well represented by an assembly of point-like electric dipole oscillators. \citet{DDA} advise that the dipole size, $\delta$, has to be chosen according to the following criterion: $|m| 2 \pi \delta / \lambda < 1/2$. This criterion is met by all grains used in our calculations and described in Sect.~\ref{sizes_and_structures}. Both \texttt{BHMIE} and \texttt{ddscat} provide the dimensionless absorption and scattering efficiencies, $Q_{abs}$ and $Q_{sca}$, respectively, and the scattering phase function $g = \langle {\rm cos}\theta \rangle$, as a function of wavelength. With $Q_{abs}$ and $Q_{sca}$, the dust mass opacity can be determined:
\begin{equation}
\kappa [{\rm cm}^2/{\rm g}] = \frac{3}{4\rho} \frac{Q_{abs} + Q_{sca}}{a},
\end{equation}
where $a$ is either the grain radius for isolated grains or the volume equivalent radius for aggregates. The single scattering albedo can also be determined: $Q_{sca}/(Q_{abs}+Q_{sca})$. When considering non-spherical particles, the optical properties have to be averaged over several grain orientations relative to the incident radiation. \citet{Mishchenko2017} made a thorough study of the conditions under which such calculations provide rigorous results. The \texttt{ddscat} routine has been constructed in order to facilitate the computation of orientational averages and meets the criteria given by \citet{Mishchenko2017} provided enough grain orientations are considered \citep[see Sect. 20 of][]{DDAmanual}. In the following, when using \texttt{ddscat}, the optical properties are averaged over 125 grain orientations as discussed in \citet{Koehler2012}, which allows relatively fast calculations and gives sufficient accuracy for our purposes.

\section{Extinction efficiency and dust mass opacity}
\label{extinction_efficiency_mass_opacity}

In this section, we explore the effects of changing the dust size, composition, and structure on the absorption and scattering efficiencies and on the dust mass opacity. A few key parameters describing these quantities will be used. First, what will be called the ``threshold wavelength'', which corresponds to the visible to near-IR wavelength above which the absorption efficiency sharply decreases. Second, variations in the spectral indices of $Q_{abs}$, $Q_{sca}$, and $\kappa$ are discussed, all of them referring to the slope of the efficiencies between 100 and 500~$\mu$m. And third, the dust mass opacities of altered grain structures are compared to that of a reference, volume-equivalent, compact sphere unless otherwise stated. The variations in the optical properties discussed in this section impact the dust temperature and far-IR emission (emissivity and spectral index).

\subsection{Isolated grains}
\label{isolated_grains_Q}

In this section we explore the size, porosity, and shape effects on the optical properties of isolated grains made of a-Sil$_{{\rm Fe,FeS}}$, a-C, and a-C:H materials. These results are then used as a reference point for aggregate dust properties.

\subsubsection{Size effects}

\begin{figure*}[!th]
\centerline{\includegraphics[width=1.4\textwidth]{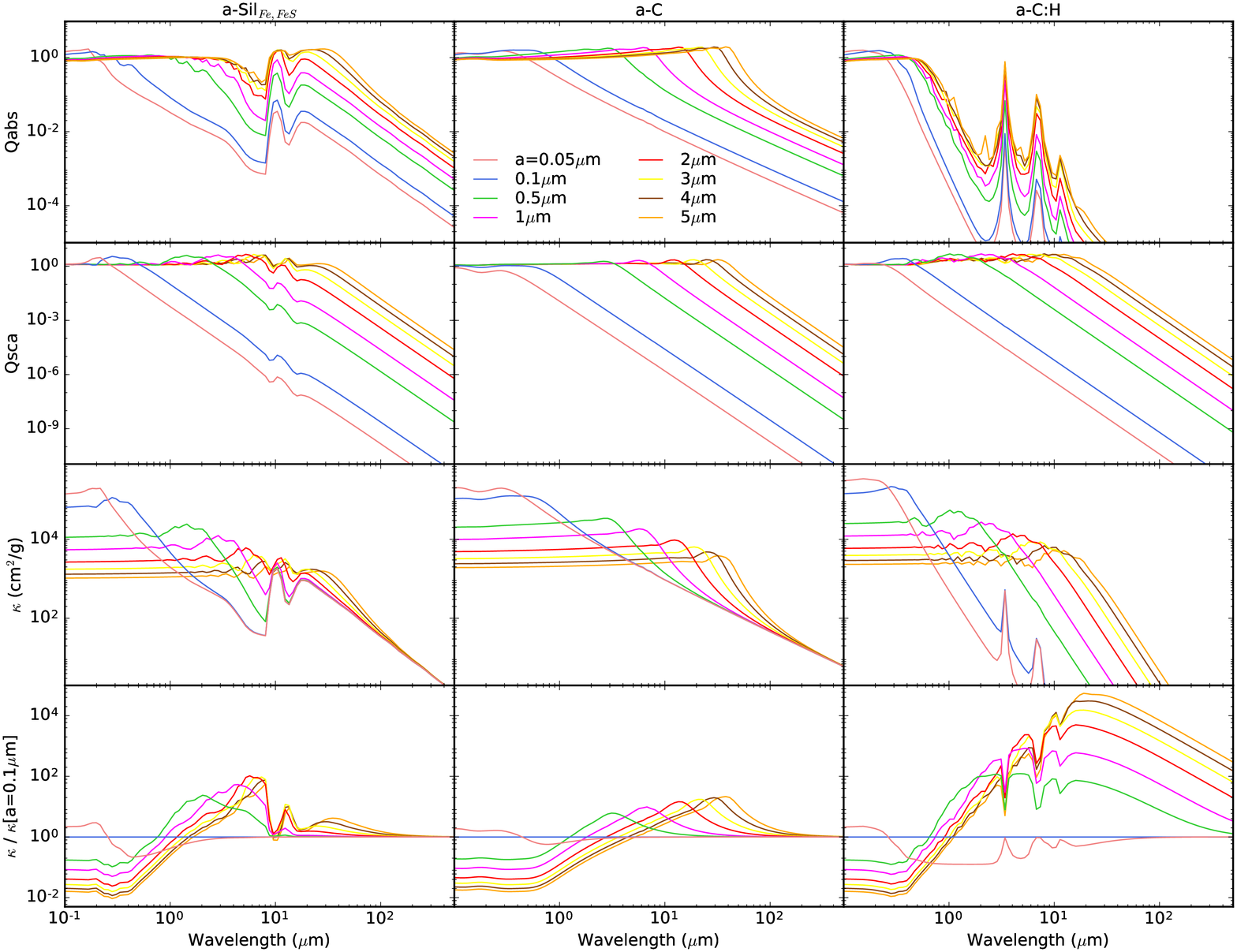}}
\caption{Influence of grain size for compact spherical a-Sil$_{{\rm Fe,FeS}}$ (left), a-C (middle), and a-C:H (right) grains with radius $a = 0.05$, 0.1, 0.5, 1, 2, 3, 4, and 5~$\mu$m plotted in pink, blue, green, fuchsia, red, yellow, brown, and orange, respectively. From top to bottom, figures show $Q_{abs}(a, \lambda)$, $Q_{sca}(a, \lambda)$, $\kappa(a, \lambda)$, and the ratio of the latter to $\kappa(a=0.1\mu$m$, \lambda)$.}
\label{Fig3} 
\end{figure*}

To explore the effect of grain size, we consider compact spheres with $a = 0.05$, 0.1, 0.5, 1, 2, 3, 4, and 5~$\mu$m (Fig.~\ref{Fig3}). The three material absorption efficiencies tend towards one at short wavelengths and decrease sharply beyond a certain threshold wavelength, which depends on both the material and the grain size. For a-C:H grains, it stagnates at 0.5~$\mu$m regardless of the size. This insulator-like behaviour is typical of high $E_g$ semi-conductors, the band gap energy of which is high enough that there is no free carrier absorption and the interband absorption occurs only at frequencies higher than the gap frequency: $E(h\nu) > E_g$(a-C:H) = 2.5~eV $\Leftrightarrow \lambda_g \sim 0.5~\mu$m. For a-C grains, the threshold wavelength varies from $\lambda \sim 2\pi a$ to $2\pi a \times 1.25$ for $0.05 \leqslant a \leqslant 5~\mu$m, independently of $E_g$(a-C) = 0.1~eV. In this case, the band gap energy is low and the grains can be considered as conductors for which free carrier absorption is important. The case of a-Sil$_{{\rm Fe,FeS}}$ grains is intermediate. Indeed, the silicates used in the optical property calculations are rather insulating but they are doped by the inclusion of conductive Fe/FeS nano-particles\footnote{If the Fe/FeS inclusions are removed from the silicates, results similar to those for a-C:H are found, with a threshold wavelength of 0.2~$\mu$m corresponding to $E_g \sim 7$~eV.}.

In the far-IR, as predicted by Mie theory, the three $Q_{abs}$ are proportional to $a$ whereas the $Q_{sca}$ are proportional to $a^4$ with $Q_{abs}$(a-Sil$_{{\rm Fe,FeS}}$) $\sim Q_{abs}$(a-C) $\sim 10^5 Q_{abs}$(a-C:H) and $Q_{sca}$(a-Sil$_{{\rm Fe,FeS}}$) $\sim Q_{sca}$(a-C) $\sim 5 Q_{sca}$(a-C:H). This agrees with the fact that insulators are not very reflective in the far-IR whereas conductors are rather bright. Fig.~\ref{Fig3} also presents the dust mass opacity $\kappa$ normalised to that of a 0.1~$\mu$m sphere. For a-Sil$_{{\rm Fe,FeS}}$ and a-C, there are almost no variations in the far-IR: $Q_{abs} \propto a \gg Q_{sca} \propto a^4$ and so $\kappa \sim Q_{abs}/a$ does not depend on the grain size. On the contrary, for a-C:H $Q_{sca} \gg Q_{abs}$ meaning that $\kappa$(a-C:H) is proportional to $a^3$ leading to strong variations.

\subsubsection{Porosity effects}

\begin{figure*}[!th]
\centerline{\includegraphics[width=1.4\textwidth]{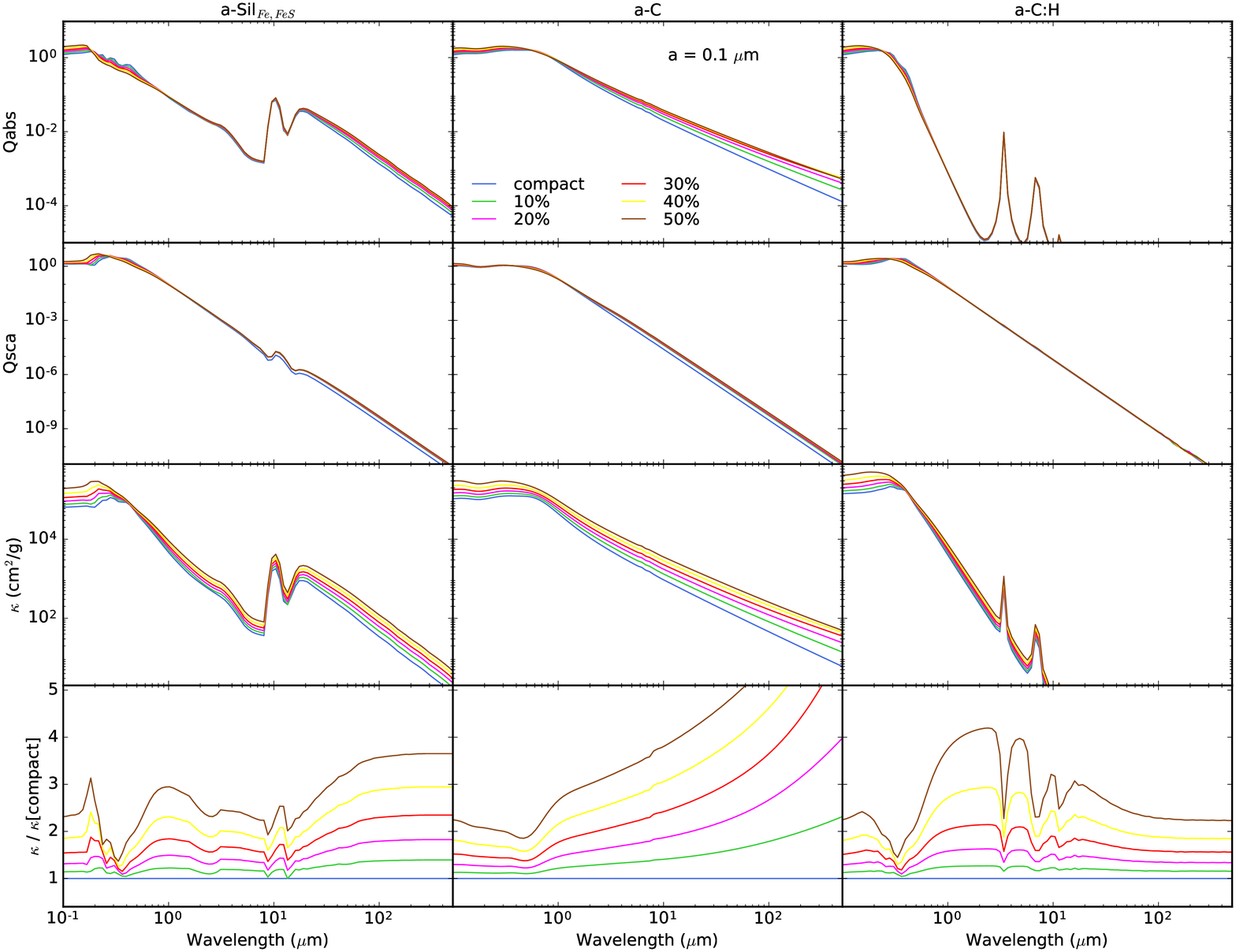}}
\caption{Influence of porosity for spherical a-Sil$_{{\rm Fe,FeS}}$ (left), a-C (middle), and a-C:H (right) grains with radius $a = 0.1 \mu$m and a porosity degree $P = 0$, 10, 20, 30, 40, and 50\% plotted in blue, green, pink, red, yellow, and brown, respectively. From top to bottom, figures show $Q_{abs}(P, \lambda)$, $Q_{sca}(P, \lambda)$, $\kappa(P, \lambda)$, and the ratio of the latter to $\kappa({\rm compact}, \lambda)$ for a compact grain with same mass.}
\label{Fig4} 
\end{figure*}

The effect of assuming porosity in a 0.1~$\mu$m spherical grain is shown in Fig.~\ref{Fig4} for $P = 0$, 10, 20, 30, 40, and 50\%. For the three materials, the threshold wavelength is shifted to shorter wavelengths, an effect already measured for semi-conductors and explained by quantum confinement in pores \citep{Vorobiev2012}. The global spectral variations of a-Sil$_{{\rm Fe,FeS}}$ and a-C:H grains are similar. For wavelengths shorter than the threshold wavelength $Q_{abs}$ increases with $P$ whereas for longer wavelengths, in the visible range, $Q_{abs}$ decreases, an effect already observed for 1~$\mu$m silicate grains by \citet{Kirchschlager2013}. The dust mass opacity increases at all wavelengths \citep{Jones1988} and we note that the far-IR spectral index does not depend on $P$. For $P >0\%$, linear relations between $P$ and the increase in the far-IR $\kappa$ can thus be found for these two materials ($\lambda \gtrsim 100~\mu$m):
\begin{eqnarray}
\frac{\kappa[{\rm porous \; a-Sil_{{\rm Fe,FeS}}]}}{\kappa[{\rm compact \; a-Sil_{{\rm Fe,FeS}}}]} \sim 5.3 P + 0.9 \\
\frac{\kappa[{\rm porous \; a-C:H]}}{\kappa[{\rm compact \; a-C:H}]} \sim 2.4 P + 0.9
\end{eqnarray}
For a-C grains, the results are quite different. First, the effect of porosity is the same as for a-Sil$_{{\rm Fe,FeS}}$ and a-C:H for wavelengths shorter than the threshold wavelength but it has almost no effect in the visible. Second, the far-IR spectral index of $Q_{abs}$, $Q_{sca}$, and $\kappa$ does depend on porosity leading to a wavelength-dependent increase. This is due to the fact that, contrary to the two other materials, the real part of the a-C refractive index $n$ is not constant in the mid- to far-IR.

\subsubsection{Shape effects}

\begin{figure*}[!th]
\centerline{\includegraphics[width=1.4\textwidth]{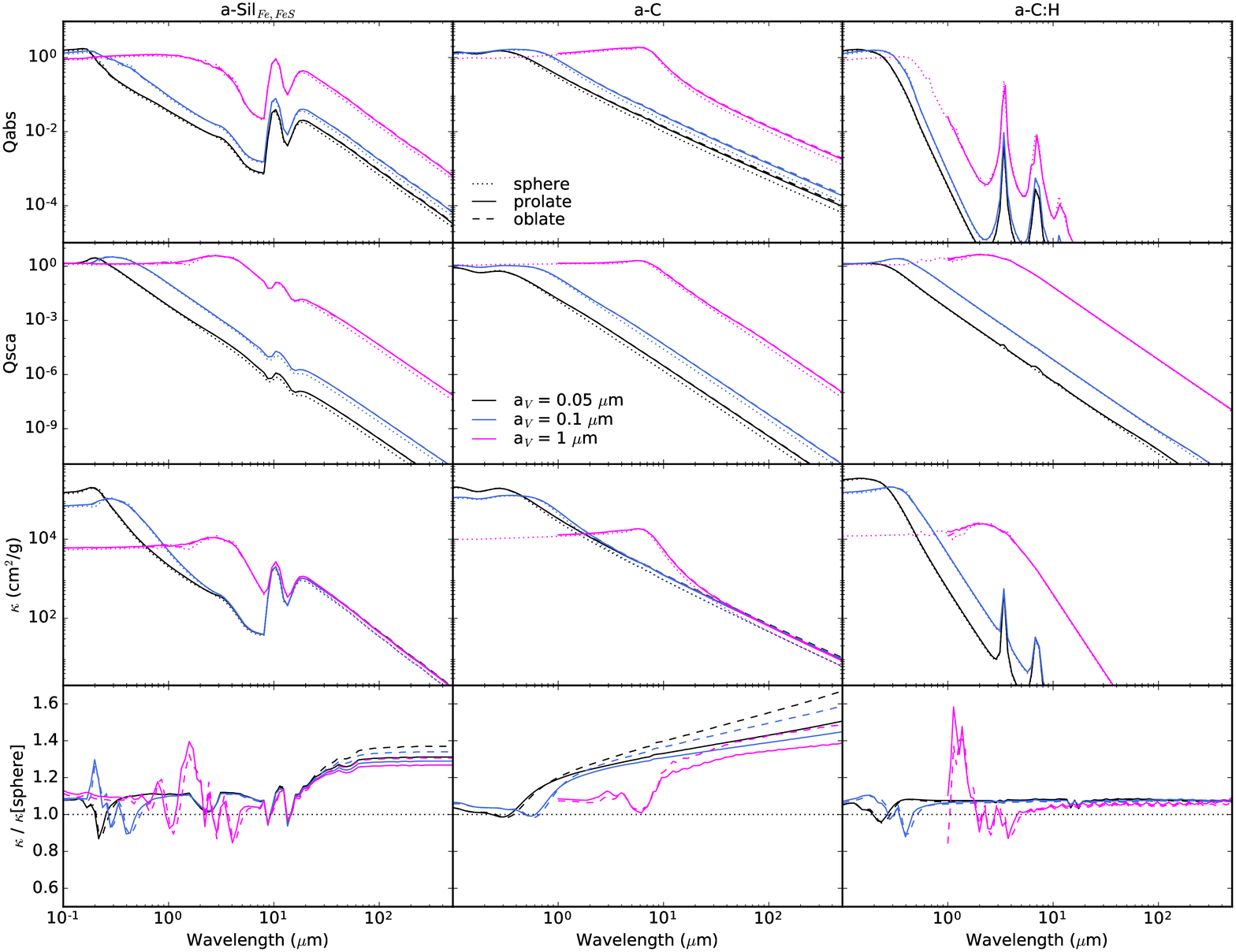}}
\caption{Influence of grain shape for compact spherical (dotted lines), oblate (dashed lines), and prolate (solid lines) a-Sil$_{{\rm Fe,FeS}}$ (left), a-C (middle), and a-C:H (right) grains with radius $a_V = 0.05$, 0.1, and 1~$\mu$m plotted in black, blue, and fuchsia, respectively. From top to bottom, figures show $Q_{abs}({\rm shape}, \lambda)$, $Q_{sca}({\rm shape}, \lambda)$, $\kappa({\rm shape}, \lambda)$, and the ratio of the latter to $\kappa({\rm sphere}, \lambda)$ for a compact spherical grain with same mass.}
\label{Fig5} 
\end{figure*}

Shape effects are shown in Fig.~\ref{Fig5}. Prolate and oblate compact grains with aspect ratio of 2 and $a_V = 0.05$, 0.1, and 1~$\mu$m are compared to compact spheres of the same mass. Whatever the material, the threshold wavelength does not depend on the grain shape. For silicates, the dust mass opacity of spheroids at $\lambda \leqslant 10~\mu$m increases by $\sim 10$\% compared to spheres and by $\sim 30$\% and $\sim 35$\% for prolate and oblate grains, respectively, at longer wavelengths. Similar results were already found in previous studies \citep[see for instance][]{Voshchinnikov2000}. Similarly, an increase is obtained for a-C grains, the main difference being the intensity of the increase and the fact that the far-IR spectral index of $Q_{abs}$, $Q_{sca}$, and $\kappa$ does depend on the shape (lower index for spheroids than for spheres) leading to a wavelength-dependent increase. In the a-C:H case the results are different: a small increase of $\sim 5$\%, which does not depend on the size or wavelength, is found. The smaller amplitude of the increase can be attributed to the insulating behaviour of a-C:H grains compared to the conductor-like behaviour of a-C and a-Sil$_{{\rm Fe,FeS}}$ grains \citep[see Chapter 12 of][for detailed explanations]{BHMIE}.

\subsection{Aggregates}
\label{aggregates_Q}

In this section, we explore the effects of the number, size, porosity, and shape of the monomers composing an aggregate.

\subsubsection{Monomer number effects}

\begin{figure*}[!th]
\centerline{\includegraphics[width=1.4\textwidth]{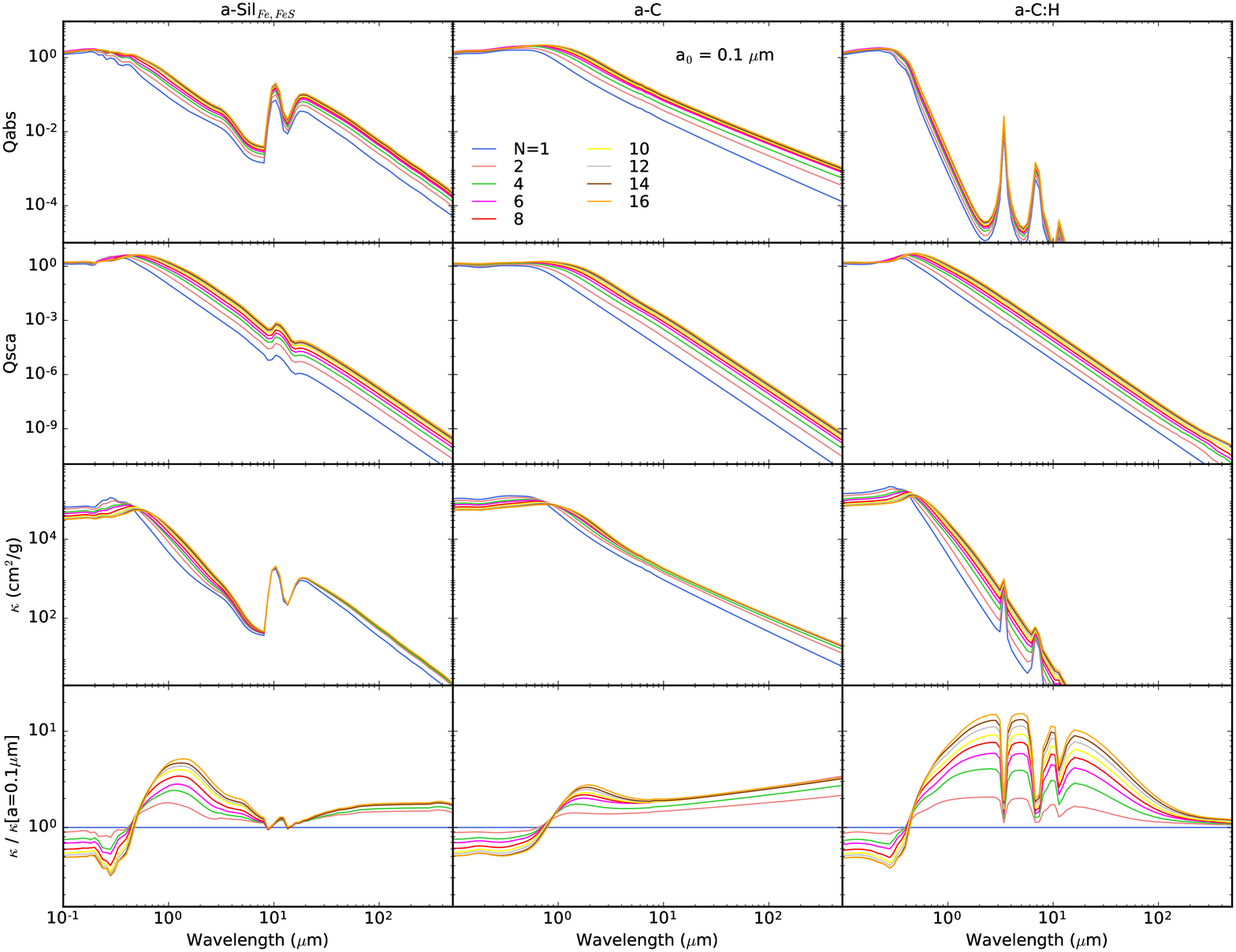}}
\caption{Influence of the number of monomers for a-Sil$_{{\rm Fe,FeS}}$ (left), a-C (middle), and a-C:H (right) aggregates made of compact spherical monomers with radius $a_0 = 0.1 \mu$m and $N = 1$, 2, 4, 6, 8, 10, 12, 14, and 16 monomers plotted in blue, pink, green, fuchsia, red, yellow, silver, brown, and orange, respectively. From top to bottom, figures show $Q_{abs}(N, \lambda)$, $Q_{sca}(N, \lambda)$, the ratio of the latter to $\kappa(N=1, \lambda)$ for an isolated compact spherical grain with $a = a_0$.}
\label{Fig6} 
\end{figure*}

Figure~\ref{Fig6} presents the optical properties of aggregates made of $2 \leqslant N \leqslant 16$ compact spherical monomers with $a_0 = 0.1~\mu$m. Behaviours already described in Sect.~\ref{isolated_grains_Q} are found. First, for a-Sil$_{{\rm Fe,FeS}}$ and a-C aggregates, the $Q_{abs}$ threshold wavelength increases with $N$ up to the threshold wavelength of the compact mass-equivalent sphere\footnote{For an aggregate composed of 0.1~$\mu$m spheres, $a_V = 0.13$, 0.16, 0.18, 0.20, 0.22, 0.23, 0.24, and 0.25~$\mu$m for $N = 2$, 4, 6, 8, 10, 12, 14, and 16, respectively.}. Second, for the insulating a-C:H aggregates the threshold wavelength stalls at $\lambda_g = 0.5~\mu$m $\Leftrightarrow E_g = 2.5$~eV as for compact spheres. However, the $\kappa$ increase in the far-IR appears to be more reminiscent of the porous spheres: indeed, the increase is at constant spectral index for a-Sil$_{{\rm Fe,FeS}}$ and a-C:H in contrast to a-C. It should be possible to draw a parallel between the ``intrinsic'' porosity of the spheres, $P$, and the ``false'' porosity induced by coagulation, $\cal{P}$. Following the definition of aggregate porosity given by \citet{Kozasa1992}, for $N = 4$, $\cal{P}$~$\sim 35$\% and for $6 \leqslant N \leqslant 16$, $\cal{P}$~$\sim 50$ to 60\%. Comparing Figs.~\ref{Fig4} and \ref{Fig6}, the increase in $\kappa$ seems much smaller for aggregates than for porous spheres. This is just the consequence of the different material densities used in the $\kappa \propto 1/\rho$ calculations: for the aggregates we use the bulk material density, whereas for porous spheres it is reduced by the porosity. So the far-IR increase seems to be dominated by induced porosity effects while the behaviour at short wavelengths appears to depend on the mass equivalent size of the aggregate. In particular, there is no blueshift of the threshold wavelength between an isolated grain and an aggregate.

\subsubsection{Monomer size effects}

\begin{figure*}[!th]
\centerline{\includegraphics[width=1.4\textwidth]{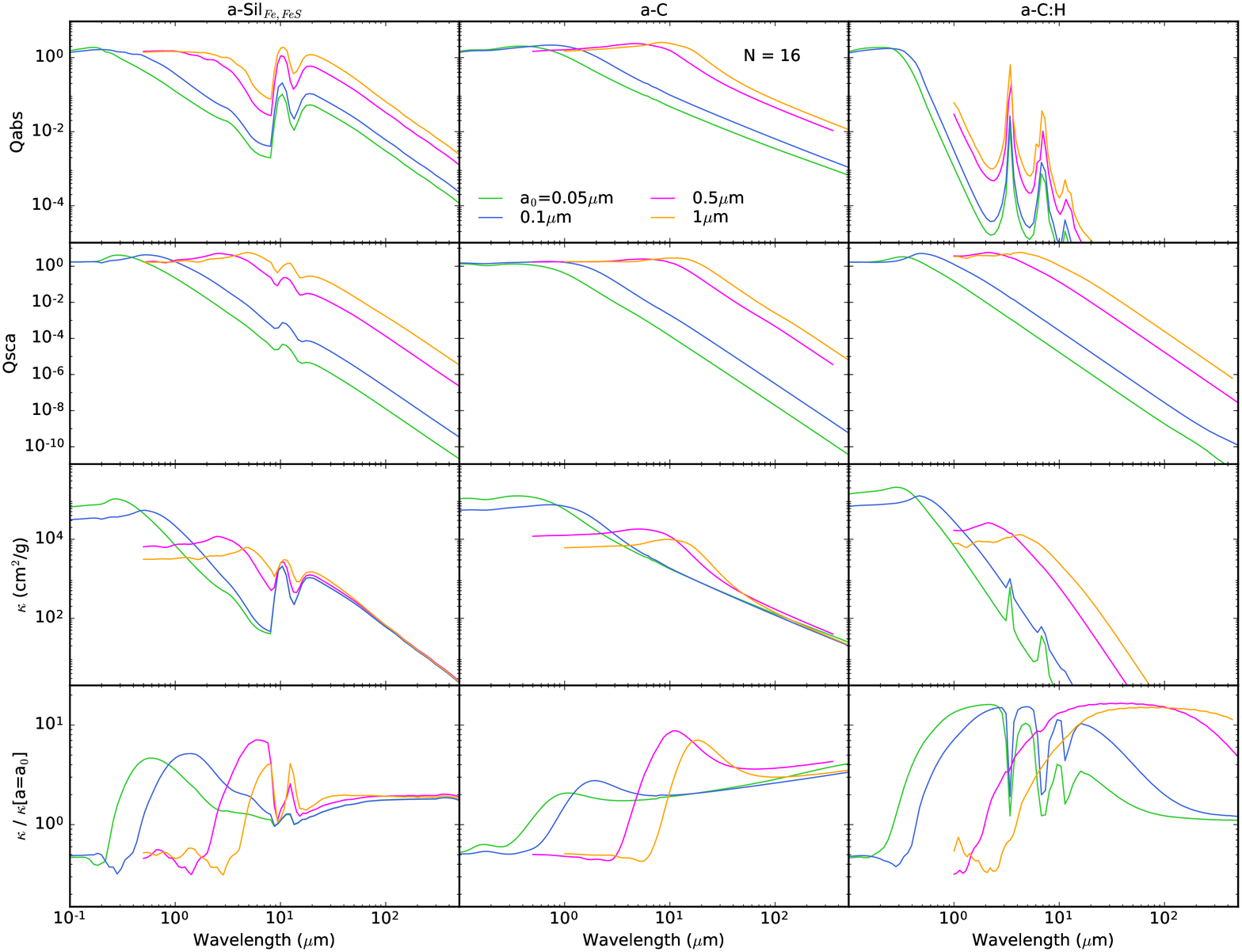}}
\caption{Influence of monomer size for a-Sil$_{{\rm Fe,FeS}}$ (left), a-C (middle), and a-C:H (right) aggregates of 16 compact spherical monomers with radius $a_0 = 0.05 \mu$m in green, 0.1~$\mu$m in blue, 0.5~$\mu$m in fuchsia, and 1~$\mu$m in orange. From top to bottom, figures show $Q_{abs}(a_0, \lambda)$, $Q_{sca}(a_0, \lambda)$, $\kappa(a_0, \lambda)$, and the ratio of the latter to $\kappa(a_0, \lambda)$ for an isolated compact spherical grain with the same radius as the monomers composing the aggregate.}
\label{Fig6bis} 
\end{figure*}

Figure~\ref{Fig6bis} shows the optical properties of aggregates made of 16 monomers with $a_0 = 0.05$, 0.1, 0.5, and 1~$\mu$m. The absorption threshold wavelength depends on the mass equivalent size of the aggregates for all monomer sizes and materials. We note that for a-Sil$_{{\rm Fe,FeS}}$ and a-C aggregates, the $\kappa$ increase in the far-IR, dominated by the increase in absorption, is a direct reflection of the porosity induced by coagulation. For a-C:H, in the case of the biggest monomers $Q_{abs} \ll Q_{sca}$ and the $\kappa$ increase becomes size-dependent.

\subsubsection{Porous monomers effects}

\begin{figure*}[!th]
\centerline{\includegraphics[width=1.4\textwidth]{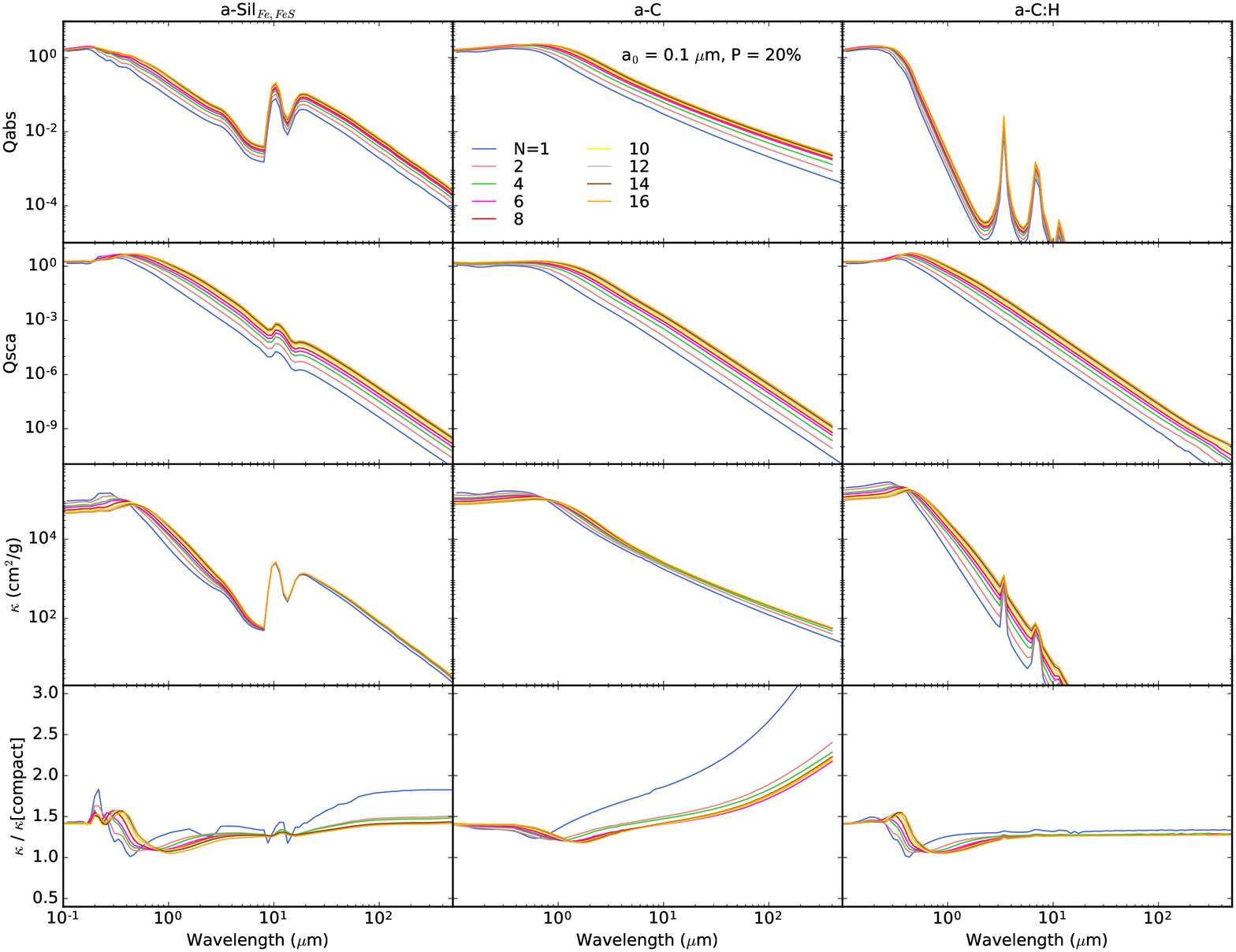}}
\caption{Influence of porosity for a-Sil$_{{\rm Fe,FeS}}$ (left), a-C (middle), and a-C:H (right) aggregates made of porous spherical monomers with radius $a_0 = 0.1 \mu$m, $P = 20$\%, and $N = 1$, 2, 4, 6, 8, 10, 12, 14, and 16 monomers plotted in blue, pink, green, fuchsia, red, yellow, silver, brown, and orange, respectively. From top to bottom, figures show $Q_{abs}(N, \lambda)$, $Q_{sca}(N, \lambda)$, $\kappa(N, \lambda)$, and the ratio of the latter to $\kappa(N, \lambda, P=0\%)$ for aggregates made of compact spherical monomers with radius $a_0 = 0.1 \mu$m.}
\label{Fig7}
\end{figure*}

Figure~\ref{Fig7} shows the optical properties of aggregates made of 0.1~$\mu$m porous monomers ($P = 20$\%) for $N = 1$ to 16 compared to aggregates of compact monomers. The results are almost the same as for porous vs. compact isolated grains with similar enhancement factors for the far-IR dust mass opacity.

\subsubsection{Monomer shape effects}

\begin{figure*}[!th]
\centerline{\includegraphics[width=1.4\textwidth]{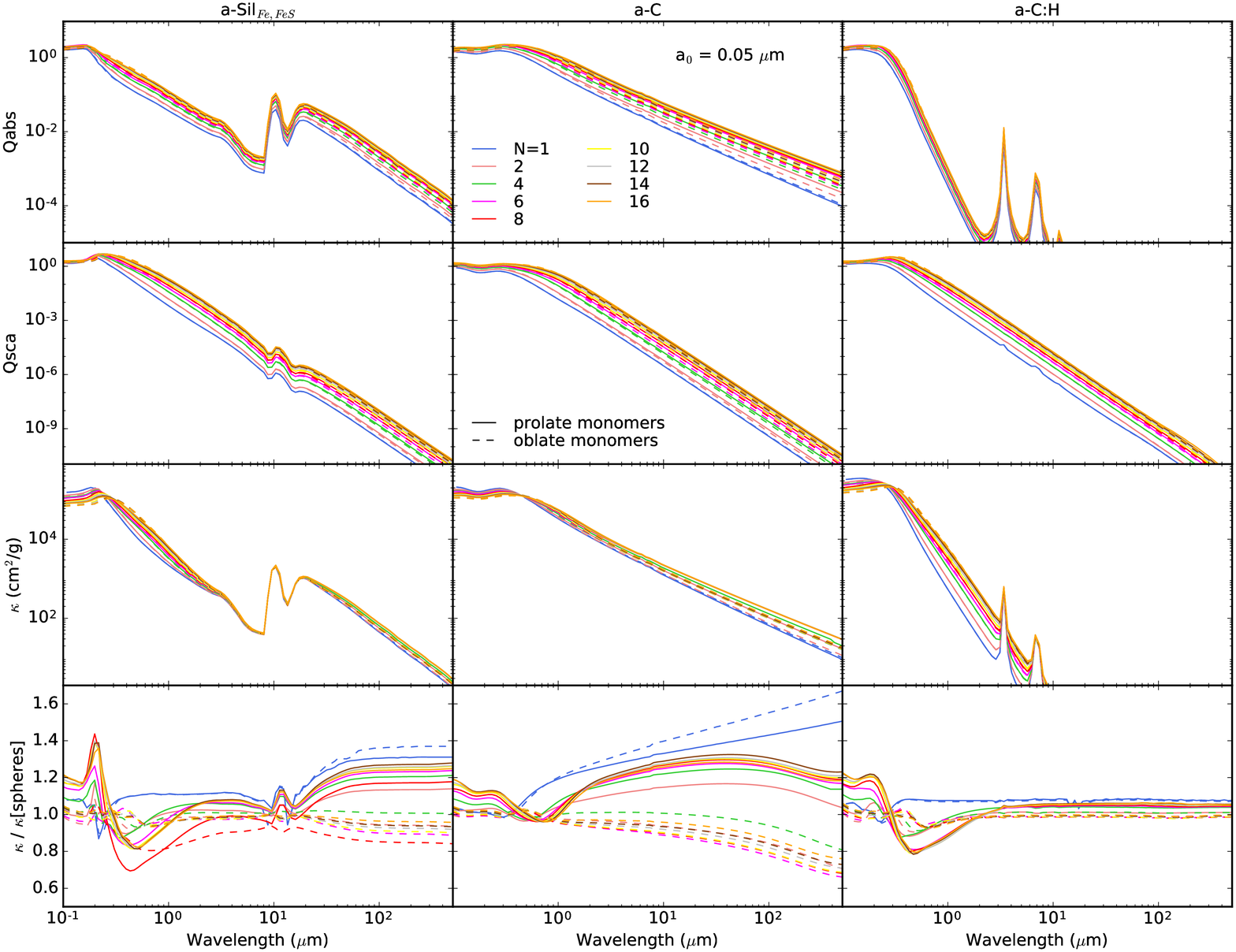}}
\caption{Influence of the monomer shape for a-Sil$_{{\rm Fe,FeS}}$ (left), a-C (middle), and a-C:H (right) aggregates made of compact spheroidal monomers with $a_V = 0.05 \mu$m, aspect ratios of 2, and $N = 1$, 2, 4, 6, 8, 10, 12, 14, and 16 monomers plotted in blue, pink, green, fuchsia, red, yellow, silver, brown, and orange, respectively. Solid lines show aggregates made of oblate monomers and dashed lines of prolate monomers. From top to bottom, figures show $Q_{abs}(N, \lambda)$, $Q_{sca}(N, \lambda)$, $\kappa(N, \lambda)$, and the ratio of the latter to $\kappa(N)$ for aggregates made of compact spherical monomers with radius $a_0 = 0.05 \mu$m.}
\label{Fig8}
\end{figure*}

The influence of monomer shapes on aggregate optical properties is presented in Fig.~\ref{Fig8}. \citet{Wu2016} have already studied the optical properties of soot aggregated with spheroidal monomers in the visible wavelength range, $0.44 \leqslant \lambda \leqslant 1~\mu$m, for $a_0 = 0.01$, 0.02, and 0.03~$\mu$m, aspect ratios from 1/3 to 3, $N = 100$, and $D_f = 1.8$ to 2.8. Differences with aggregates made of spherical monomers were found: an increase in both $Q_{abs}$ and $Q_{sca}$ by $\sim 5$\% for aspect ratios of 1/2 and 2. We confirm this trend in the visible for both oblate and prolate monomers. 

At longer wavelengths, $Q_{abs}$ and $Q_{sca}$ are higher for aggregates made of prolate monomers than oblates. However when comparing the dust mass opacities of aggregates of spheroids and aggregates of spheres, strong variations are found. First, for aggregates made of prolate monomers, an increase is found in the far-IR (see also Sect.~\ref{far_IR_opacity} and Fig.~\ref{Fig17} for a detailed discussion about $\kappa$ variations at 250~$\mu$m). This increase can be explained by two factors. As seen in the previous section, isolated prolate grains have higher $\kappa$ in the far-IR than isolated spherical grains so it not surprising to again find this characteristic when the grains are aggregated. Then, aggregates made of prolate monomers have higher ``false'' porosities than aggregates made of spheres: $\cal{P} =$ 40 to 50\% for spheres and 50 to 60\% for prolates. We also notice that the increase between aggregates of spheres and aggregates of prolates is lower than the increase between isolated spheres and prolates. Indeed, on average, aggregates have a smaller eccentricity than isolated prolates and it has been shown that this difference increases with eccentricity \citep[e.g.][]{Voshchinnikov2000}. Second, for aggregates made of oblate monomers, a decrease is found in the far-IR compared to aggregates of spheres (see also Sect.~\ref{far_IR_opacity} and Fig.~\ref{Fig17}), even though they exhibit high $\cal{P}$ values from 70 to 80\%. Following \citet{Kozasa1992}, a surface-area equivalent radius can be estimated as $a_S = \sqrt{\cal{A}/\pi}$, where $\cal{A}$ is defined as the projected area of the aggregate averaged over three orthogonal directions. Compared to aggregates of spheres, $a_S$ for aggregates of prolate monomers are $\sim$ 10\% smaller while they are $\sim$ 30\% bigger for aggregates of oblate monomers. Since $\kappa$ is proportional to $Q_{ext}$ which is itself proportional to $1/a_S^2$, this explains why aggregates of oblates have lower dust mass opacities. Nonetheless the big differences due to variations in monomer shapes should be mitigated. Indeed, since in our aggregates the monomer principal axes are all aligned (see Sect.~\ref{sizes_and_structures}), and all the monomers have the same axis ratio, the variations emphasised here represent upper limits of the effect of having non-spherical monomers in the aggregates.

A last interesting point concerns the spectral index of $\kappa$ in the far-IR for a-C grains, which is dominated by absorption. As seen in Figs.~\ref{Fig6} and \ref{Fig6bis}, the coagulation of spheres leads to a flattening of $\kappa$. Fig~\ref{Fig8} shows that this also remains true for spheroids but that this flattening is weaker. For $100 \leqslant \lambda \leqslant 200~\mu$m, the difference in slopes is $\sim$ 0.05 and 0.1 for a-C prolate and oblate monomers, respectively.

\subsubsection{Mixed monomer size effects}

\begin{figure*}[!th]
\centerline{\includegraphics[width=1.4\textwidth]{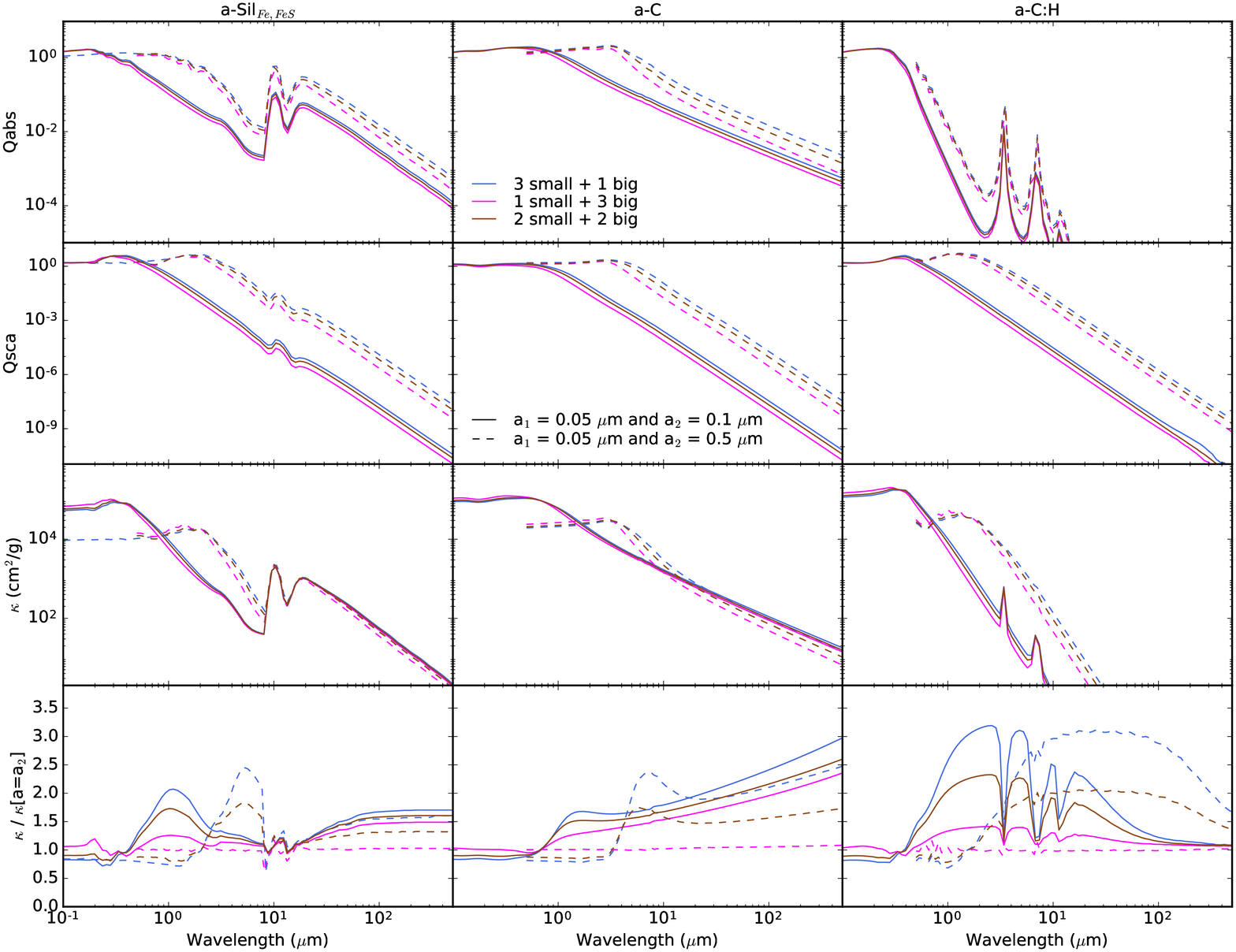}}
\caption{a-Sil$_{{\rm Fe,FeS}}$ (left), a-C (middle), and a-C:H (right) aggregates made of compact spherical monomers with radius $a_1 = 0.05 \mu$m and $a_2 = 0.1$ (solid lines) or 0.5~$\mu$m (dashed lines), and $N = 4$. Plotted in blue are the results for aggregates made of three grains with $a_0 = a_1$ and one grain with $a_0 = a_2$, in fuchsia one grain with $a_0 = a_1$ and three grains with $a_0 = a_2$, and in green two grains of each radius. From top to bottom, figures show $Q_{abs}(a_2, \lambda)$, $Q_{sca}(a_2, \lambda)$, $\kappa(a_2, \lambda)$, and the ratio of the latter to $\kappa(a = a_2)$ for an isolated compact spherical grain with radius $a = a_2$.}
\label{Fig9}
\end{figure*}

Finally, Fig.~\ref{Fig9} shows the effect of building an aggregate out of monomers with different radii. A similar study was already undertaken by \citet{Liu2015} in which soot aggregates composed of 50 to 500 monomers with $a_0 \sim 33 (45) \pm 1.3 (1.1)$~nm with $D_f = 1.8$ were considered. The calculations were made in the visible ($\lambda = 0.65~\mu$m) and showed that, even if the monomers had very close sizes, both the scattering and absorption efficiencies were higher than in the case of aggregates made of monomers of same size with the same total mass. \citet{Koehler2012} considered the case of aggregates composed of one big $a_0 = 60$~nm monomer surrounded by 250 to 2000 smaller monomers with $a_0 = 3.5$~nm. Their results were similar to those of \citet{Liu2015} with an increase in the absorption efficiency both in the visible and far-IR wavelength ranges. Fig.~\ref{Fig9} presents the case where there are only four monomers but of larger sizes: $a_1 = 0.05~\mu$m and $a_2 = 0.1$ or 0.5~$\mu$m. In the $a_2/a_1 = 10$ case, the smallest monomers always represent less than 1\% of the total aggregate volume for all the structures considered and in the $a_2/a_1 = 2$ case, they represent 4 to 27\% of the volume, proportions close to those in \citet{Koehler2012} and \citet{Liu2015}. As in the two previous studies, both $Q_{abs}$ and $Q_{sca}$ are increased and the increase is larger when more small monomers are incorporated into the aggregate which can be likened to an increase of the monomer surface irregularity. This result holds for all three of the aggregate compositions considered. 

\subsection{Far-IR dust mass opacity}
\label{far_IR_opacity}

\begin{figure*}[!th]
\centerline{\includegraphics[width=1.4\textwidth]{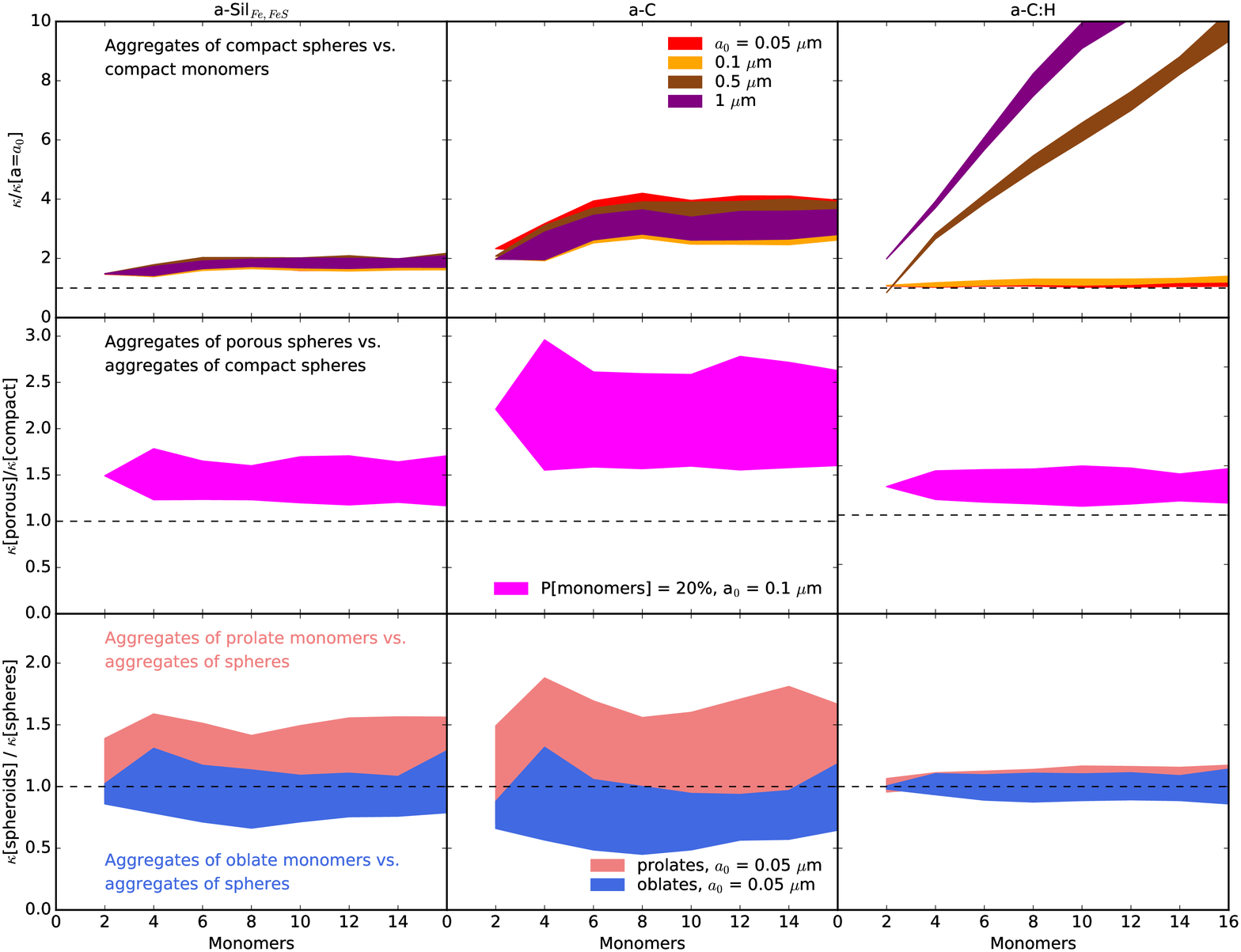}}
\caption{a-Sil$_{{\rm Fe,FeS}}$ (left), a-C (middle), and a-C:H (right) aggregate normalised dust mass opacities at 250~$\mu$m as a function of the number of monomers composing the aggregates, $N$. Top row shows $\kappa(a_0, N)$ for $a_0 = 0.05$ (red), 0.1 (orange), 0.5 (brown), and 1~$\mu$m (purple) normalised to $\kappa(a_0, N=1)$. Middle row shows $\kappa(a_0 = 0.1~\mu{\rm m}, N, P = 20\%)$ normalised to $\kappa(a_0 = 0.1~\mu{\rm m}, N, P = 0\%)$. Bottom row shows $\kappa$(oblate monomers, $N$) in blue and $\kappa$(prolate monomers, $N$) in pink normalised to $\kappa$(spherical monomers, $N$). The filled areas show the scatter of the dust mass opacities for 10 randomly chosen aggregate shapes (see Sect.~\ref{sizes_and_structures}).}
\label{Fig17}
\end{figure*}

A far-IR dust mass opacity (emissivity) enhancement is observed towards many dense molecular clouds where dust growth is expected \citep[][among many others]{Remy2017, Suutarinen2013, Roy2013, Rawlings2013, Martin2012}. Mass opacity estimates from models are also commonly used to derive dust masses and temperatures from astronomical far-IR data. As $\kappa$ depends on the dust composition, size, and structure, we investigate the influence of these three parameters. Fig.~\ref{Fig17} shows dust mass opacities at 250~$\mu$m for aggregates made of compact and porous spherical monomers, and of oblate and prolate compact monomers, as a function of the number of monomers.

For aggregates made of compact spheres, we see that for a-Sil$_{{\rm Fe,FeS}}$ and a-C grains, the increase in $\kappa$ does not depend on the monomer size $a_0$ but only on the number of monomers $N$. The surface to volume ratio of the aggregates can be estimated by $a_S/a_V$ and is found to depend only on $N$, not on $a_0$. Then, similarly to \citet{Koehler2011}, we find that the increase in $\kappa$ for $N \geqslant 6-8$ saturates at $\sim 1.8 \pm 0.2$ for a-Sil$_{{\rm Fe,FeS}}$ grains and $\sim 3.3 \pm 0.7$ for a-C grains, which means that the aggregate total opacity is not proportional to the sum of the monomer opacities. The threshold at $N \sim 6-8$ matches the monomer number above which $a_S < a_V$, and thus the saturation is explained by the shadowing effects of the aggregate cross-sections as described in \citet{Jones2011}. The ``intensity'' of the shadowing is roughly proportional to the number of monomers. In the case of aggregates made of a-C:H grains, the results are similar for $a_0 = 0.05$ and 0.1~$\mu$m with a $\kappa$ increase $\sim 1.1 \pm 0.1$ for $N \geqslant 6-8$. However, for larger monomer radius, a strong increase in the far-IR $\kappa$ is found which is due to the subordination of the absorption to the scattering efficiency as already shown in Fig.~\ref{Fig6bis}.

When porous monomers are considered, the behaviour of the increase in $\kappa$ as a function of $N$ is similar to the case of compact monomers consistent with the behaviour for $N>6$ observed in Fig.~\ref{Fig17} for aggregates made of porous vs. compact monomers. The $\kappa$ increase is $\sim 1.45 \pm 0.25$, $\sim 2.1 \pm 0.5$, and $\sim 1.3 \pm 0.2$, for a-Sil$_{{\rm Fe,FeS}}$, a-C, and a-C:H aggregates, respectively. This matches the increase found between isolated compact and porous spheres presented in Fig.~\ref{Fig4}.

For aggregates made of compact spheroids compared to aggregates made of compact spheres, as shown in Sect.~\ref{aggregates_Q}, the variations depend on the shape of the monomers. For aggregates made of prolate monomers, we find an increase of $\sim 1.3 \pm 0.1$, $\sim 1.3 \pm 0.2$, and $\sim 1.05 \pm 0.05$ for a-Sil$_{{\rm Fe,FeS}}$, a-C, and a-C:H grains, respectively. For aggregates made of oblate monomers, a decrease of $\sim 0.8 \pm 0.1$, $\sim 0.8 \pm 0.1$, and $1.00 \pm 0.02$ is found for a-Sil$_{{\rm Fe,FeS}}$, a-C, and a-C:H grains, respectively. Consequently, if the aggregates were built out of a distribution of spheroids and not only prolate or oblate monomers, on average their far-IR dust mass opacities would be close to that of an aggregate made of spheres in the far-IR.

\section{Silicate mid-IR spectral features}
\label{silicate_midIR_feature}

\begin{figure*}[!th]
\centerline{\includegraphics[width=1.4\textwidth]{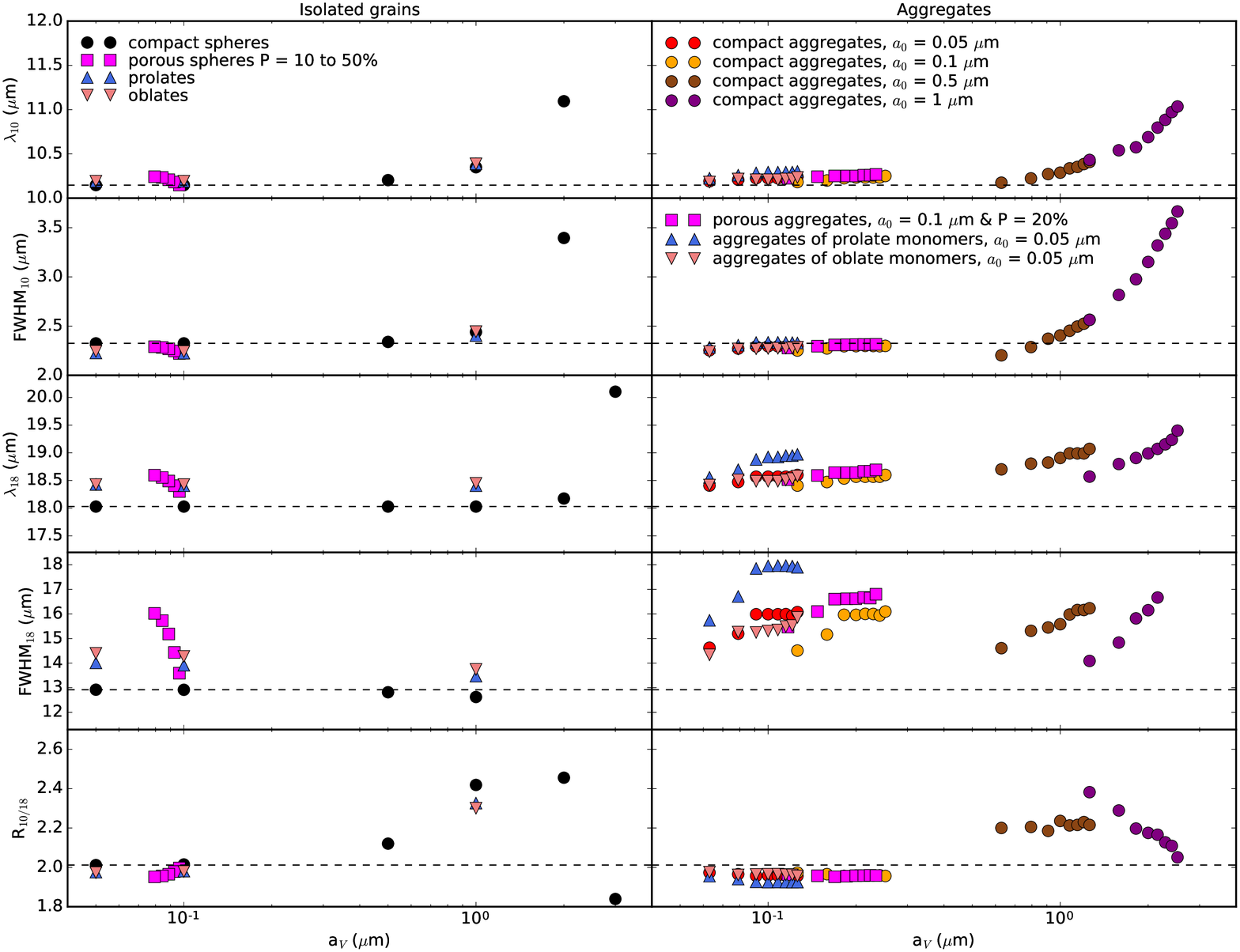}}
\caption{Peak positions, $\lambda_{10}$ and $\lambda_{18}$ (first and third rows), full widths at half maximum, FWHM$_{10}$ and FWHM$_{18}$ (second and fourth rows), and ratios of the dust mass opacities of the mid-IR silicate features, R$_{10/18}$ (bottom row), for isolated grains (left colum) and aggregates (right column). $\lambda$, FWHM, and R$_{10/18}$ are plotted as a function of the radius of a compact sphere with same mass for compact spheres (black circles), porous spheres (fuchsia squares), prolate grains (blue triangles), oblate grains (coral triangles), and aggregates of compact spheres with $a_0 = 0.05$ (red circles), 0.1 (orange circles), 0.5 (brown circles), 1~$\mu$m (purple circles), aggregates of porous spheres with $P = 20\%$ and $a_0 = 0.1~\mu$m (fuchsia squares), aggregates of prolate monomers with $a_0 = 0.05~\mu$m (blue triangles), and aggregates of oblate monomers with $a_0 = 0.05~\mu$m (coral triangles). The horizontal black dashed lines show $\lambda$, FWHM, and R$_{10/18}$ of the smallest isolated compact sphere.}
\label{Fig18}
\end{figure*}

The amorphous silicate mid-IR spectral features (10~$\mu$m Si-O stretching mode and 18~$\mu$m O-Si-O bending mode) are often used as a diagnostic to determine the dust composition and size in a variety of astronomical environments, ranging from the diffuse ISM to the Galactic centre, as well as from AGB star envelopes to protoplanetary disk \citep[see for instance the review by][]{Henning2010}. First, if $a_V \geqslant 1~\mu$m, the contrast between the continuum and the two features decreases strongly as can be seen in Figs.~\ref{Fig3} and \ref{Fig6bis}, independently of the detailed grain structure (aggregate or isolated grain). Consequently, it would be very hard to detect and thus to use observations of the silicate bands in absorption to prove the presence of grains larger than 2~$\mu$m in dense media. However, the detailed study of the peak positions, shapes and ratio of the two bands may be used to constrain the population of grains with $a \sim 0.5-1~\mu$m, together with complementary observations of light scattering by grains at shorter wavelengths. Second, Fig.~\ref{Fig18} shows the peak positions, $\lambda_{10}$ and $\lambda_{18}$, and the full widths at half maximum, FWHM$_{10}$ and FWHM$_{18}$, of the 10~$\mu$m and 18~$\mu$m silicate extinction features\footnote{Both the peak position and full with at half maximum are calculated directly from the unnormalised dust mass opacity. Contrary to what is usually done when analysing astonomical observations, we do not subtract any continuum nor normalise the curves. As a result, FWHM cannot be defined when the features become very wide, explaining the ``missing points'' in Fig.~\ref{Fig18} for large $a_V$.}, respectively, for the isolated grains and aggregates described in this study. Also shown are the ratios of the $\kappa$ values at the peak positions of the two features, R$_{10/18} = \kappa(\lambda_{10})/\kappa(\lambda_{18})$. Both $\lambda$ and FWHM depend strongly on the grain structure as already shown by various authors with different descriptions of grain irregularities: distributions of hollow spheres, gaussian random field particles, porous grains, spheroidal grains, spherical/spheroidal grains with increasing sizes \citep[][among others]{Bouwman2001, Kemper2004, Min2007, Breemen2011}.

For isolated spherical grains, both $\lambda_{10}$ and FWHM$_{10}$ are almost constant as long as $a < 1~\mu$m and increase for larger particles. Indeed, above $a \sim 1~\mu$m, the extinction band has a significant contribution from the scattering ($Q_{sca} \sim Q_{abs}$), whereas at smaller sizes the band profile depends only on the absorption. For non-spherical grains, the feature is red-shifted with a slightly smaller FWHM$_{10}$. A red-shift of the feature with increasing porosity is also observed at constant FWHM$_{10}$. Interestingly, because the aggregate far-IR $\kappa$ can be related to the aggregate ``false'' porosity (Sect.~\ref{aggregates_Q}), so can the increase in $\lambda_{10}$ of aggregates relative to $\lambda_{10}$ for the mass equivalent spheres. Moreover, as for isolated grains, the aggregate $\lambda_{10}$ and FWHM$_{10}$ are almost constant as long as the aggregate mass equivalent sizes are smaller than 1~$\mu$m but increase for larger sizes. However, both increases are smaller in the case of aggregates than in the case of isolated grains with $\Delta\lambda_{10} = -0.4~\mu$m and $\Delta$FWHM$_{10} = -0.25~\mu$m for $a_V = 2~\mu$m. The shape of the monomers composing the aggregates has little influence on the 10~$\mu$m feature shape with a shift smaller than $\sim 0.05~\mu$m of the peak position for all $a_V$ and no variations in FWHM$_{10}$.

Variations in $\lambda_{18}$ and FWHM$_{18}$ are appreciably different. For isolated spherical grains, $\lambda_{18}$ increases when $a > 1~\mu$m but FWHM$_{18}$ decreases for $a \geqslant 0.5~\mu$m with a shift $\Delta$FWHM$_{18} \sim -0.3~\mu$m from $a = 0.1$ to 1~$\mu$m. Contrary to the 10~$\mu$m feature, the scattering contribution to the 18~$\mu$m extinction feature remains negligible even at $a = 1~\mu$m. A decrease is also found for spheroids with $\Delta$FWHM$_{18} \sim -0.5~\mu$m for $a_V = 0.05$ to 1~$\mu$m. As for the 10~$\mu$m feature, a red-shift and a broadening of the 18~$\mu$m feature with increasing porosity is observed. However, the variations are stronger with $\Delta\lambda_{18} \sim 0.6~\mu$m and $\Delta$FWHM$_{18} \sim 3~\mu$m for $P = 0$ to 50\%. For aggregates, there is a clear difference in the variations of the two features, the strongest one being the dependence of both $\lambda_{18}$ and FWHM$_{18}$ on the monomer size $a_0$ rather than on the mass equivalent size $a_V$. For $a_0 = 0.05$ and 0.1~$\mu$m, the peak position $\lambda_{18}$ does not vary significantly and can be related to the aggregate ``false'' porosity as can be $\lambda_{10}$. But when larger monomers are considered, the increase in $\lambda_{18}$ is no longer monotonous with $a_V$. The dependence of FWHM$_{18}$ on the monomer size is conspicuous for all $a_0$ considered. Monomer shapes also influence the feature: when compared to aggregates of spheres, in the case of prolate monomers $\Delta\lambda_{18} \sim 0.4~\mu$m and $\Delta$FWHM$_{18} \sim 2~\mu$m for $a_V = 0.1~\mu$m but $\Delta\lambda_{18} \sim 0~\mu$m and $\Delta$FWHM$_{18} \sim -0.7~\mu$m in the case of oblate monomers for the same $a_V$. The O-Si-O 18~$\mu$m bending mode is thus more dependent on the grain detailed structure than the Si-O 10~$\mu$m stretching mode, both in peak position and width/spectral shape (see Figs.~\ref{Fig6}, \ref{Fig6bis}, and \ref{Fig8}).

For isolated grains, the ratios of the $\kappa$ values at the peak positions of the two mid-IR features, R$_{10/18}$, are almost constant for the smallest grains: variations of less than 5\% are found for $0.05 \leqslant a \leqslant 0.5~\mu$m, less than 3\% when porosity is introduced ($P = 10$ to 50\%), and less than 2\% when oblate and prolate grains are considered instead of spheres. For larger grains, the variations are stronger and not monotonous. For compact spheres with $a = 1$, 2, and 3~$\mu$m, increases of 17\%, 18\%, and 9\% are found, respectively, with respect to a 0.05~$\mu$m sphere. The results are similar for aggregates: R$_{10/18}$ varies only by $\sim 3$ to 5\%, in comparison with a 0.05~$\mu$m sphere, for monomers with $a_0 = 0.05$ and 0.1~$\mu$m, porous or compact, spherical or spheroidal. Significant variations appear for larger monomers: an increase of $\sim 10$\% for $a_0 = 0.5~\mu$m spherical monomers, and for $a_0 = 1~\mu$m, an increase up to 15\% when $N = 2$ and down to 2\% for $N = 16$. As $\lambda_{18}$ and FWHM$_{18}$, R$_{10/18}$ thus depends on the monomer size. The increase in R$_{10/18}$ for large $a_V$ comes from the increasing contribution of scattering to the 10~$\mu$m extinction feature. The decreasing ratios found with increasing $a_V$ and $a$ for the largest aggregates and isolated grains considered are due to the fact that for these sizes the 20~$\mu$m feature keeps on increasing whereas the 10~$\mu$m feature saturates.

\section{Albedo and asymmetry factor}
\label{scattering_properties}

One way of probing grain growth in dense interstellar media is the measure of scattered light from the visible to the near- and mid-IR, known as cloudshine and coreshine. The amount of scattered light reaching an observer depends, at least, on the dust albedo and phase function. In this section, we focus on the grain scattering properties, showing single scattering albedos, $Q_{sca}/Q_{ext}$, and the asymmetry factors of the phase function, $g = <{\rm cos}\,\theta>$, for isolated grains and aggregates of a-Sil$_{{\rm Fe,FeS}}$, a-C, and a-C:H.

\begin{figure*}[!th]
\centerline{\includegraphics[width=1.4\textwidth]{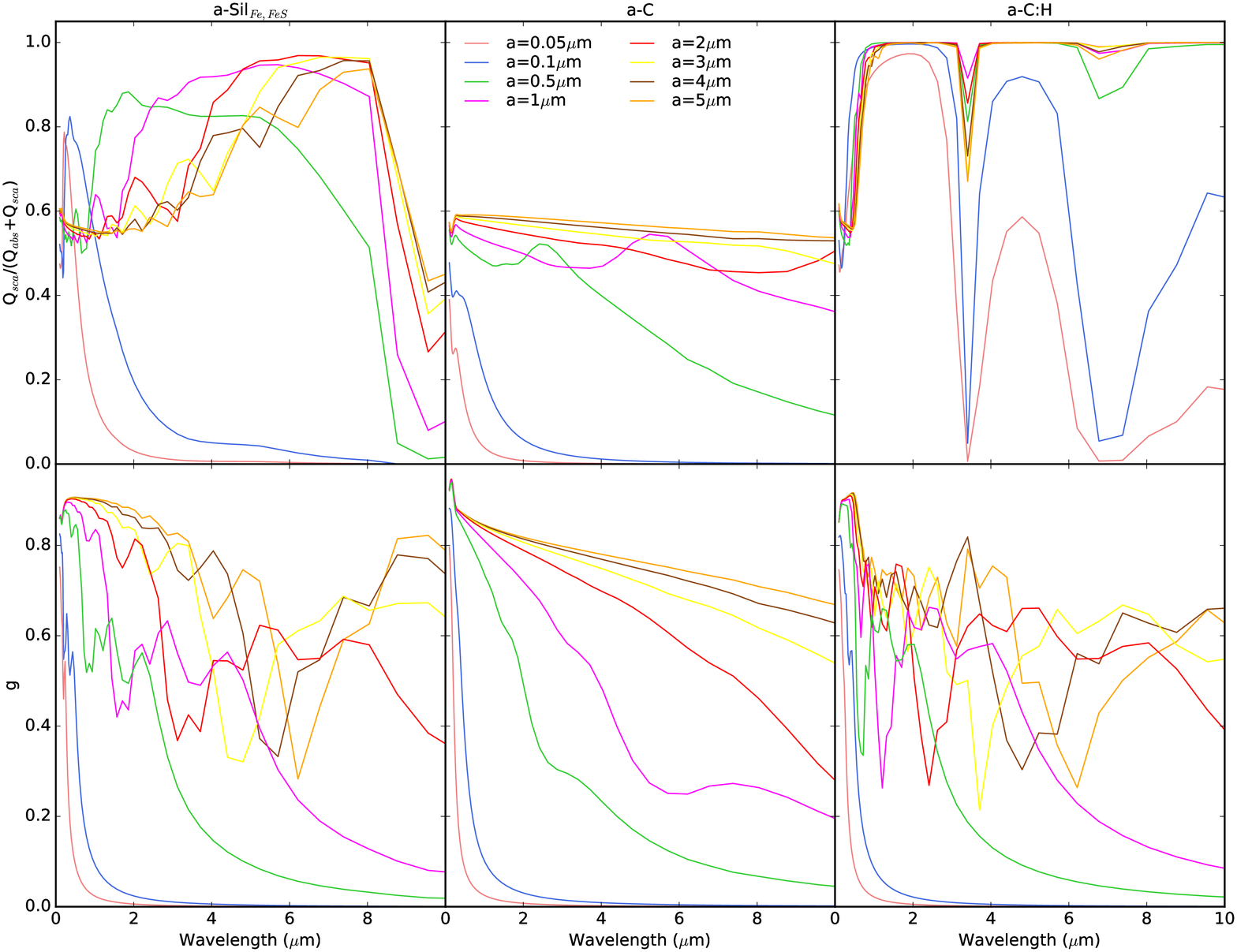}}
\caption{Influence of size for compact spherical a-Sil$_{{\rm Fe,FeS}}$ (left), a-C (middle), and a-C:H (right) grains with radius $a = 0.05$, 0.1, 0.5, 1, 2, 3, 4, and 5~$\mu$m plotted in pink, blue, green, fuchsia, red, yellow, brown, and orange, respectively. Top: albedo $Q_{sca}/(Q_{abs}+Q_{sca})$. Bottom: asymmetry factor of the phase function $g = <{\rm cos}\,\theta>$.}
\label{Fig10} 
\end{figure*}

\begin{figure*}[!th]
\centerline{\includegraphics[width=1.4\textwidth]{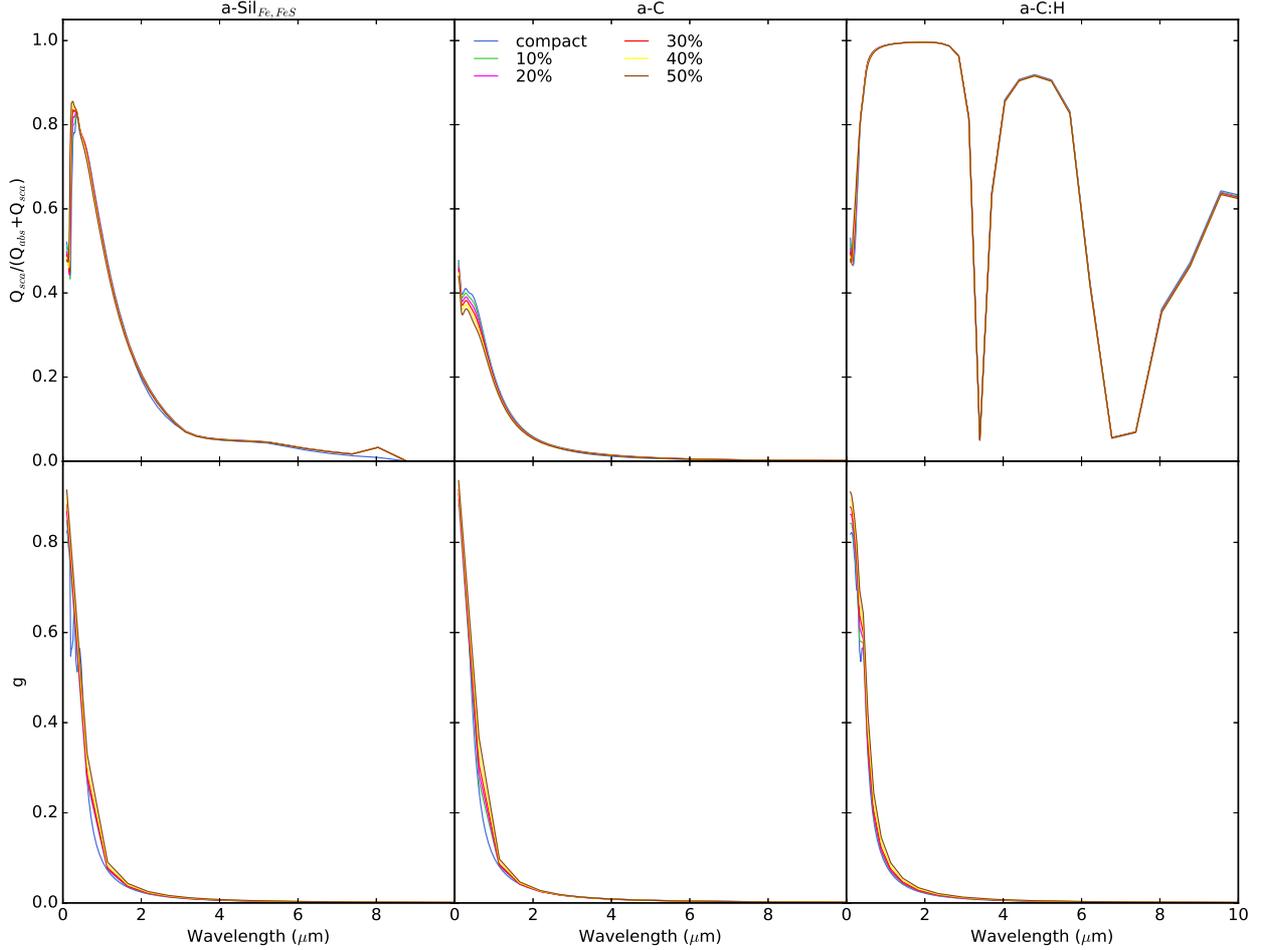}}
\caption{Influence of porosity for spherical a-Sil$_{{\rm Fe,FeS}}$ (left), a-C (middle), and a-C:H (right) grains with radius $a = 0.1 \mu$m and porosity degree $P = 10$, 20, 30, 40, and 50\% plotted in blue, green, pink, red, yellow, and brown, respectively. Top: $Q_{sca}/(Q_{abs}+Q_{sca})$. Bottom: $g = <{\rm cos}\,\theta>$.}
\label{Fig11} 
\end{figure*}

\begin{figure*}[!th]
\centerline{\includegraphics[width=1.4\textwidth]{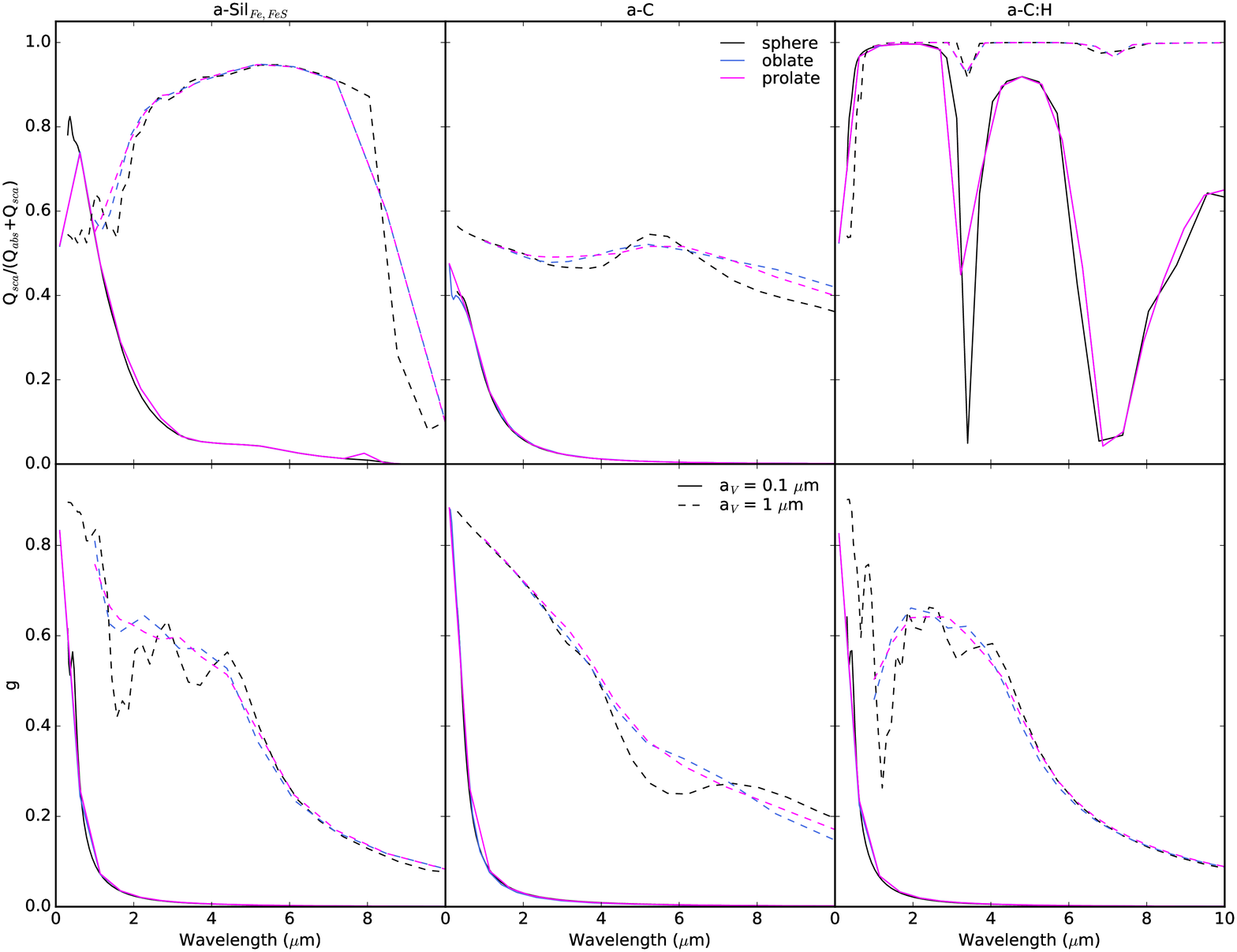}}
\caption{Influence of shape for spherical (black), oblate (blue), and prolate (magenta) a-Sil$_{{\rm Fe,FeS}}$ (left), a-C (middle), and a-C:H (right) grains with radius $a_V = 0.1$ and 1~$\mu$m, solid and dashed lines, respectively. Top: $Q_{sca}/(Q_{abs}+Q_{sca})$. Bottom: $g = <{\rm cos}\,\theta>$. The asymmetry factor curves appear smoother because the calculations were made for lesser wavelength points.}
\label{Fig12} 
\end{figure*}

Fig.~\ref{Fig10} shows the scattering properties of compact spherical grains. For a-Sil$_{{\rm Fe,FeS}}$ grains, at very short wavelengths the albedo is about 0.5, reflecting the fact that absorption and scattering efficiencies are each close to one. A sharp increase in the albedo occurs at a given wavelength corresponding to the first interference peak position in the scattering efficiency (``bump'' in the $Q_{sca}$ plots related to interference between incident light and backward-scattered light). This wavelength depends on both grain size and composition. For the a-Sil$_{{\rm Fe,FeS}}$ grains, it goes from $\sim 0.2$ to 7~$\mu$m for $0.05 \leqslant a \leqslant 5~\mu$m. For $a \geqslant 0.5~\mu$m, the albedo remains rather high before decreasing due to absorption in the 10~$\mu$m band. For a-C grains, absorption and scattering have similar amplitudes and no spectral features from the near- to mid-IR for $a \geqslant 0.5~\mu$m so the albedo stays around 0.5 over the entire wavelength range. For a-C:H grains, the extinction is dominated by scattering and the albedo is equal to 1 except at the mid-IR band positions in the 3.3-3.4~$\mu$m and 7-9~$\mu$m regions. As detailed in \citet[][see their Sect.~2.2]{Jones2016}, the subordination of absorption to scattering can be explained by Anderson localisation of absorbing electrons which occurs in disordered (amorphous) materials \citep{Anderson1958}.

For the three materials, the asymmetry parameter variations with grain size are classical: isotropic scattering for the smallest grains and dominant forward scattering for the biggest for $\lambda \geqslant 1~\mu$m. The effect of porosity for isolated grains is very small with only a 0.1~$\mu$m blue-shift of the sharp increase in albedo (Fig.~\ref{Fig11}), which matches the shift observed in the absorption efficiency when $P$ goes from 0 to 50\% (Fig.~\ref{Fig4}). Fig.~\ref{Fig12} shows albedos and asymmetry factors for spheroids, which are very similar to those of spheres. This is consistent with the findings of \citet{Mishchenko1996} for low-$k$ materials and extends it to highly absorbing materials such as a-C. Our results also agree with \citet{Voshchinnikov2000} who showed that the difference in albedo between spheres and spheroids remains small in the visible as long as $k/n \gtrsim 0.2-0.3$, even for aspect ratios up to 10.

\begin{figure*}[!th]
\centerline{\includegraphics[width=1.4\textwidth]{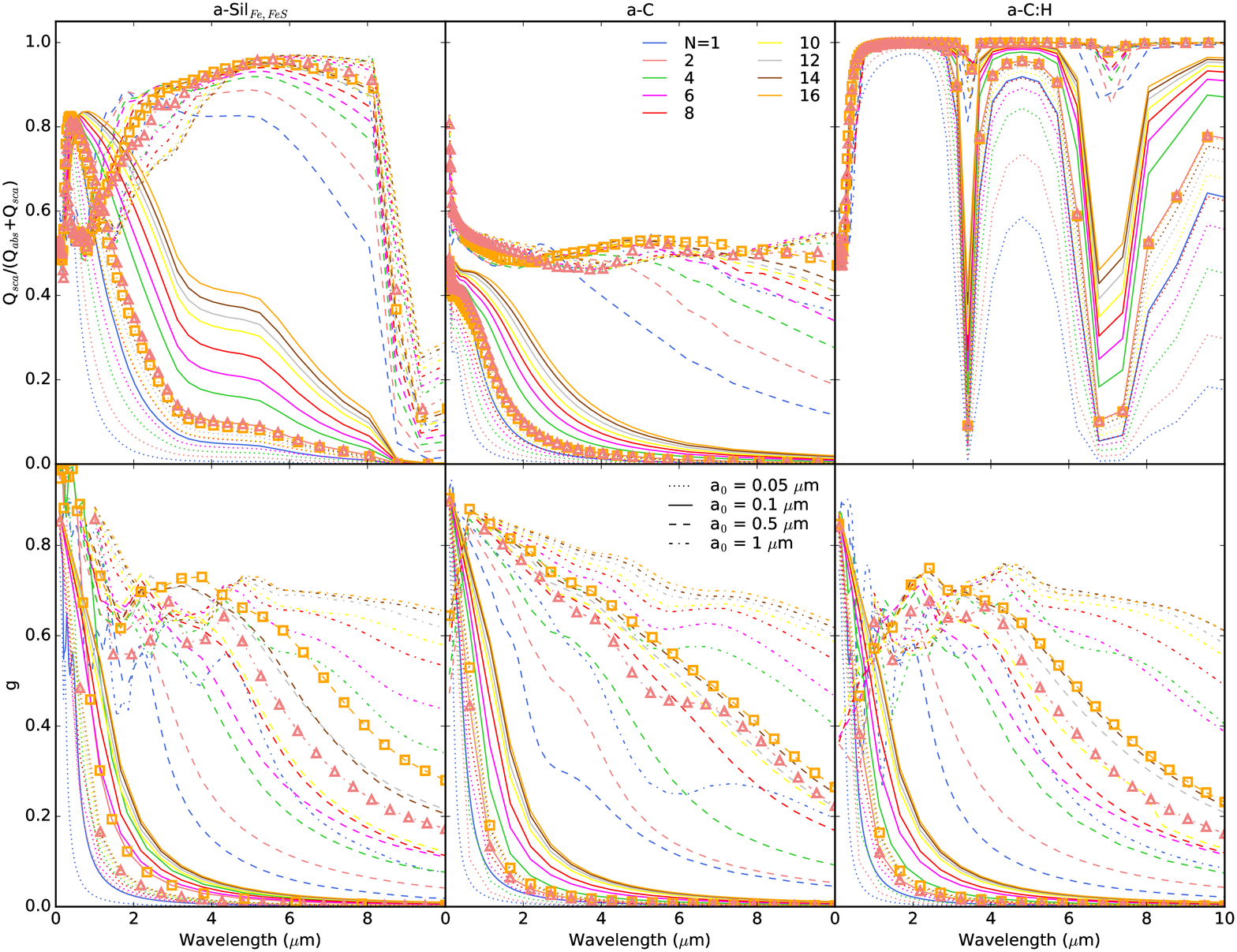}}
\caption{Influence of monomer size and number for a-Sil$_{{\rm Fe,FeS}}$ (left), a-C (middle), and a-C:H (right) aggregates made of compact spherical monomers with radius $a_0 = 0.05$ (dotted lines), 0.1 (solid lines), 0.5 (dashed lines), and 1~$\mu$m (dot-dashed lines) and $N = 1$, 2, 4, 6, 8, 10, 12, 14, and 16 monomers plotted in blue, pink, green, fuchsia, red, yellow, silver, brown, and orange, respectively. Top: $Q_{sca}/(Q_{abs}+Q_{sca})$. Bottom: $g = <{\rm cos}\,\theta>$. Aggregates with $N = 16$ and $a_0 = 0.05$ and 0.5~$\mu$m are also highlighted with triangles, while aggregates with $N = 2$ and $a_0 = 0.1$ and 1~$\mu$m are highlighted with squares.}
\label{Fig13} 
\end{figure*}

\begin{figure*}[!th]
\centerline{\includegraphics[width=1.4\textwidth]{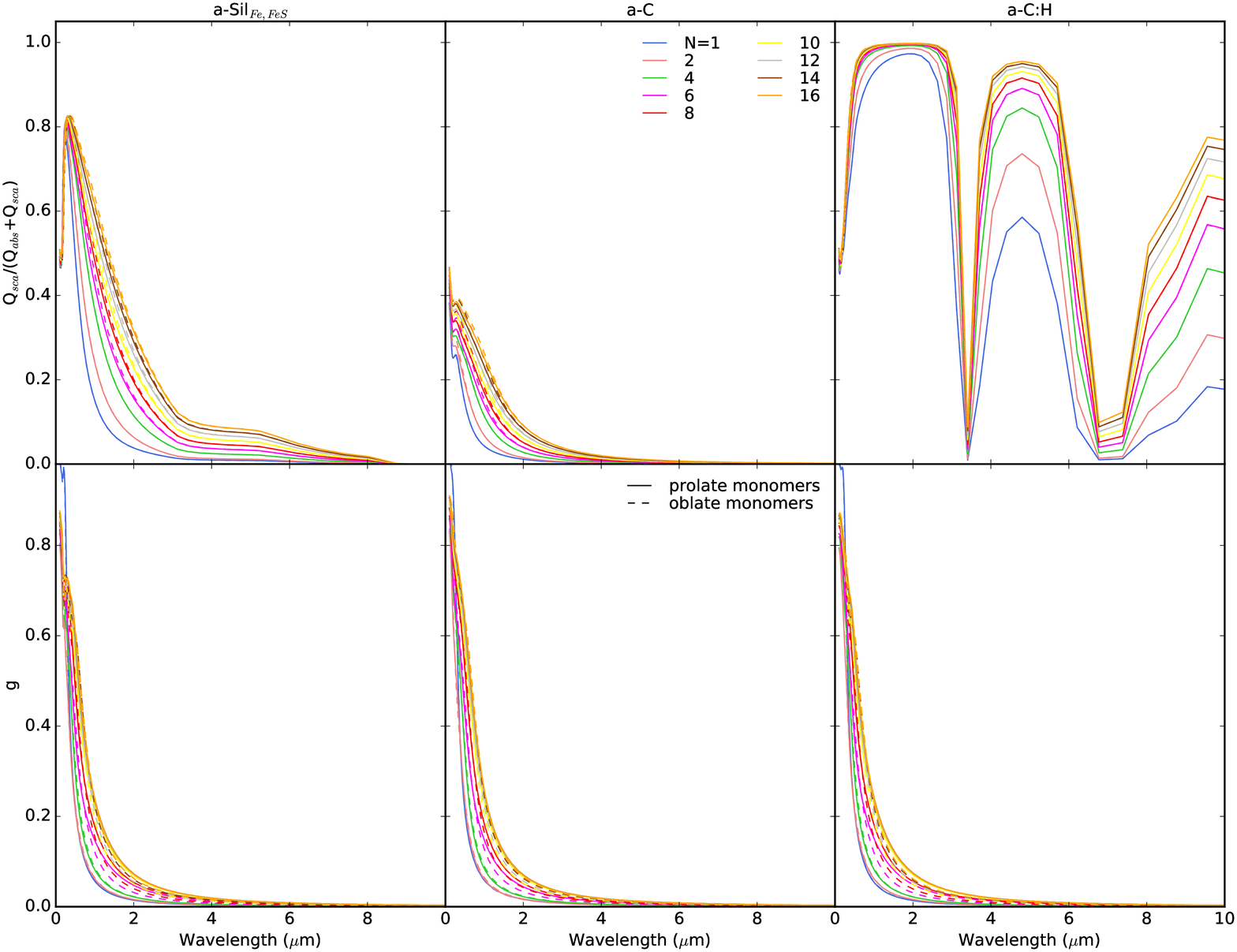}}
\caption{Inlfuence of monomer shape for a-Sil$_{{\rm Fe,FeS}}$ (left), a-C (middle), and a-C:H (right) aggregates made of compact spheroidal monomers with radius $a_0 = 0.05 \mu$m, aspect ratios of 2, and $N = 1$, 2, 4, 6, 8, 10, 12, 14, and 16 monomers plotted in blue, pink, green, fuchsia, red, yellow, silver, brown, and orange, respectively. Solid lines show aggregates made of oblate monomers, dashed lines of prolate monomers, and dotted lines made of spherical monomers. Top: $Q_{sca}/(Q_{abs}+Q_{sca})$. Bottom: $g = <{\rm cos}\,\theta>$.}
\label{Fig15} 
\end{figure*}

The scattering properties of aggregates made of compact spheres are shown in Fig.~\ref{Fig13}. Aggregate albedos are almost exactly the same as albedos of compact spheres with same mass and do not depend on the monomer size. For instance, aggregates made of two 0.1 (1)~$\mu$m monomers and of sixteen 0.05 (0.5)~$\mu$m monomers have the same mass, and thus the same $a_V$, and exhibit very similar albedos (see orange squares and pink triangles, respectively, in Fig.~\ref{Fig13}). On the contrary, the asymmetry factor of the phase function depends on the monomer size: two aggregates of same mass but with different monomer radius have a $\sim 0.1$ difference in $g$ values in favour of more forward scattering for the aggregate with the smallest monomers. Fig.~\ref{Fig15} shows albedos and asymmetry factors for aggregates made of compact spheroids. From the visible to the near-IR, the albedos of aggregates made of spherical, oblate, and prolate monomers do not differ by more than 5\%. Variations in the asymmetry factors are bigger with an increase of $\sim$~5\% and 10\% at $\lambda = 0.55~\mu$m for aggregates of prolate and oblate monomers, respectively, and of $\sim$~25\% and 10\% at $\lambda = 1.63~\mu$m. Our results agree with the results of \citet{Wu2016} for $D_f = 2.8$ at $\lambda = 0.87~\mu$m (green curves in their Fig.~7).

\section{DDA vs. approximate light scattering models}
\label{DDA_vs_EMT}

\begin{figure*}[!th]
\centerline{
\begin{tabular}{c}
\includegraphics[width=1.15\textwidth]{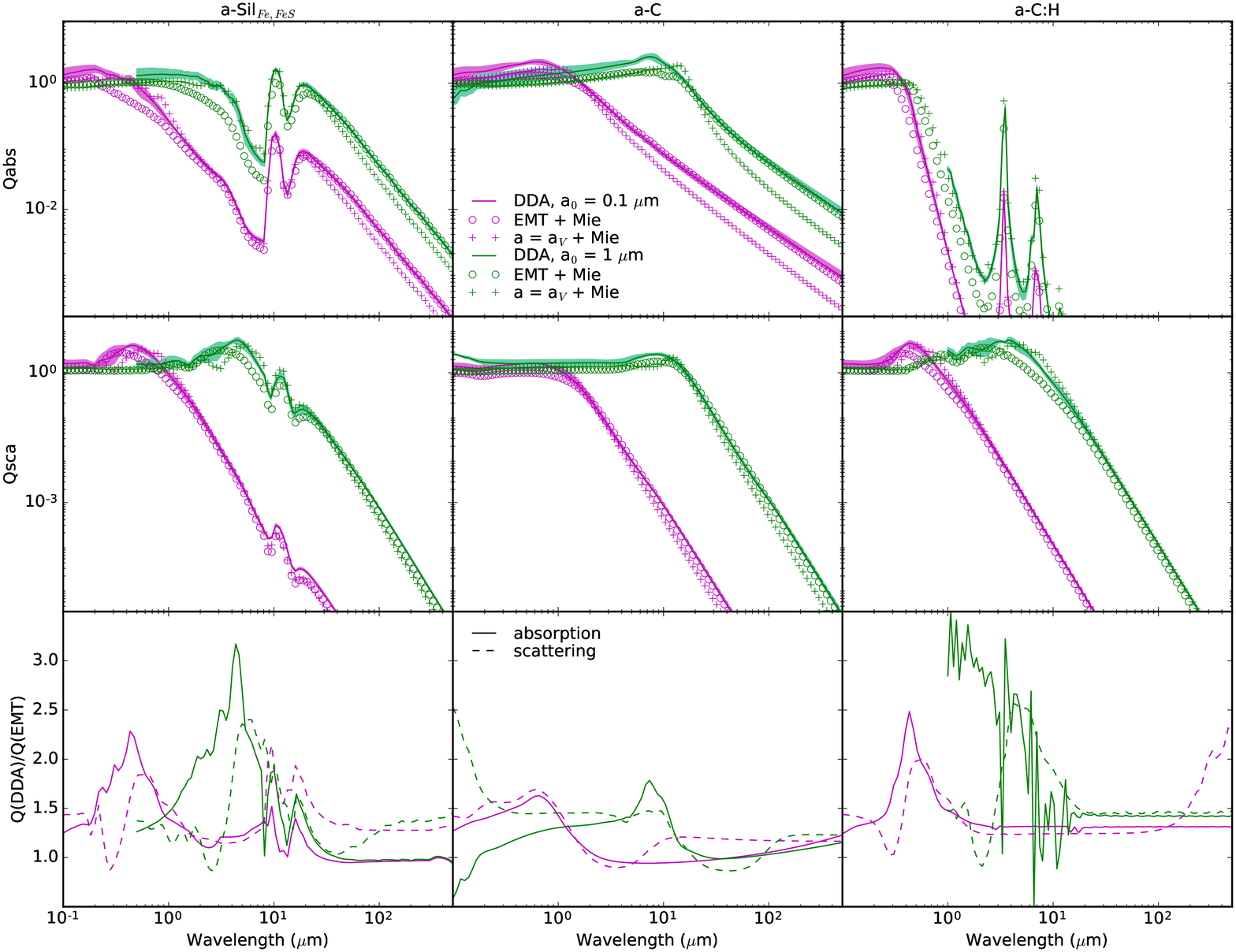} \\
\includegraphics[width=1.15\textwidth]{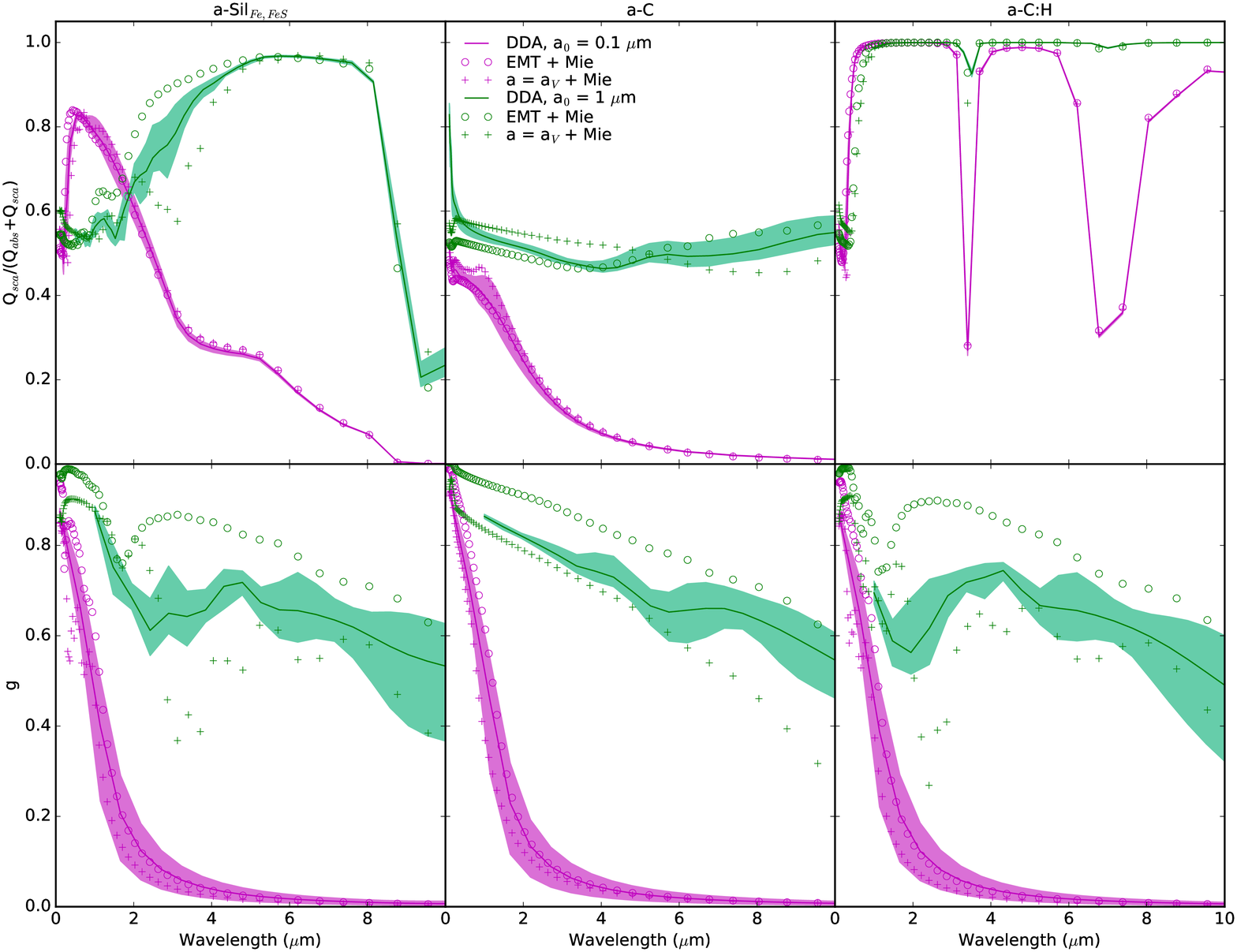}
\end{tabular}}
\caption{Comparison of the optical properties of an aggregate made of 8 monomers calculated with DDA and with an equivalent using Mie theory using EMT (see Sect.~\ref{DDA_vs_EMT} for details) for a-Sil$_{{\rm Fe,FeS}}$ (left), a-C (middle), and a-C:H (right) grains. From top to bottom, the rows show $Q_{abs}$, $Q_{sca}$, the ratio of the two previous quantities calculated with DDA vs. EMT, the albedo, and the asymmetry factor of the phase function. The solid lines are the DDA optical properties for $a_0 = 0.1$ (magenta) and 1~$\mu$m (green). The filled areas show the scatter of the dust properties for 10 randomly chosen aggregate shapes (see Sect.~\ref{sizes_and_structures}). The crosses show the Mie calculations for the compact sphere of equivalent mass ($a = a_V$). The empty circles show the Mie calculations using optical constants obtained through EMT.}
\label{Fig19} 
\end{figure*}

The DDA method is very convenient thanks to its high flexibility in terms of grain geometry and composition but computation time can be quite long. It is thus always convenient to be able to minimise at least part of the calculations using approximate methods such as an effective medium theory (EMT). \citet{Mishchenko2016b, Mishchenko2016a} already showed that EMT can provide accurate estimates of the scattering matrix and absorption cross-section of heterogeneous grains (i.e. grains with inclusions or porous grains) if two criteria are met: both the size parameter of the inclusions and the refractive index contrast between the host material and the inclusions have to be small. In this section, we compare our results for aggregates with results obtained for spheres with average dielectric functions estimated using the Maxwell Garnett (MG) mixing rules \citep{BHMIE, MG1904} and Mie theory, refered to as MG+Mie in the following\footnote{The computations were done using the publicly available {\it emc} routine of V. Ossenkopf: \url{https://hera.ph1.uni-koeln.de/~ossk/Jena/pubcodes.html}.}. We assume that a grain consists of a sphere of a-Sil$_{{\rm Fe,FeS}}$, a-C, or a-C:H, with spherical vacuum inclusions. The MG sphere radius, $a_C$, is defined as the radius of the homogeneous sphere with the same radius of gyration, $R_g$, as the corresponding aggregate \citep{Kozasa1992}:
\begin{eqnarray}
a_C &=& \sqrt{\frac{5}{3}} R_{g} {\rm ,}\\
R_g^2 &=& \frac{1}{N} \sum (\overrightarrow{r_i} - \overrightarrow{r_0})^2{\rm ,}
\end{eqnarray}
where $\overrightarrow{r_0}$ is the position vector of the aggregate mass centre and $\overrightarrow{r_i}$ is the position vector of the $i$th monomer centre. The vacuum inclusion fraction or porosity can then be defined as:
\begin{equation}
\cal{P} = {\rm 1} - {\it N\left(\frac{a_0}{a_C}\right)}^{\rm 3}.
\end{equation}
Fig.~\ref{Fig19} shows the example of MG+Mie calculations with two aggregates made of 8 compact spherical monomers with $a_0 = 0.1$ and 1~$\mu$m calculates using DDA. The MG+Mie spheres have $a_C = 0.28$ and 2.9~$\mu$m and $\cal{P}$ = 60 and 65\%, respectively. Also shown are the optical properties for compact spheres with the same mass, with $a_V = 0.2$ and 2~$\mu$m, respectively, which reproduce correctly both the threshold wavelength and the shape of the 10~$\mu$m silicate band but are not good analogues otherwise \citep[similar results in e.g.][]{Tazaki2018}.

For a-Sil$_{{\rm Fe,FeS}}$ aggregates, the MG+Mie grains reproduce almost perfectly the far-IR $Q_{abs}$ but underestimate $Q_{sca}$ by 30 to 40\% in this same wavelength range. Similar discrepancies, up to a few tens of percent depending on the wavelength range, were found by \citet{Min2016} for aggregates made of spherical monomers with $a_V = 0.2$ to 4~$\mu$m. The MG+Mie grains also fail to reproduce the mid-IR spectral features which was expected according to Fig.~\ref{Fig18} and the threshold wavelengths which are blue-shifted. This shift due to the inclusion of porosity in the MG+Mie grain matches the shifts observed in Fig.~\ref{Fig4} for isolated porous grains. This results in an underestimate of $Q_{abs}$ by factors greater than 1.5 for $0.2 \leqslant \lambda \leqslant 1~\mu$m and $a_0 = 0.1~\mu$m and greater than 2 for $2 \leqslant \lambda \leqslant 8~\mu$m and $a_0 = 1~\mu$m. The scattering efficiency is also underestimated even if to a lesser extent. The MG+Mie grain albedos are close to aggregate albedos in the mid-IR when $a_0 = 0.1~\mu$m but overestimated by $\sim 20$\% when $a_0 = 1~\mu$m. Similarly, for the smallest monomer size, the asymmetry parameter is well estimated by MG+Mie but for bigger monomers, it is overestimated by $\sim 40$\% in the 2 to 4~$\mu$m wavelength range, and by 15 to 20\% at longer mid-IR wavelengths. For a-C and a-C:H aggregates, the discrepancies are similar except for the far-IR $Q_{abs}$, which is not as well reproduced by MG+Mie as for a-Sil$_{{\rm Fe,FeS}}$ grains and is underestimated by 10 to 50\%, and we even note differences in the spectral index for a-C grains. Differences in the spectral index can be explained by the inclusion of porosity as described in Sect.~\ref{isolated_grains_Q} (Fig.~\ref{Fig4}).

Over the years, many approximate methods have been developed and tested against numerically exact calculations such as DDA or the T-matrix method \citep[TMM,][]{TMM}. A thorough comparison of many of these methods was recently undertaken by \citet{Tazaki2018} for silicate aggregates. Seven usual approximate methods were confronted: the Rayleigh-Gans-Debye theory \citep{Tazaki2016}, the mean field theory \citep{Berry1986, Botet1997}, the MG+Mie theory \citep{Mukai1992, Kataoka2014}, the Mie theory with aggregate polarisability mixing rules \citep[APMR Mie,][]{Min2008}, the MG+Mie with the simple percolation model \citep{Henning1996}, the distribution of hollow spheres \citep[DHS,][]{Min2016, Jones1988}, and the continuous distribution of ellipsoids \citep[CDE,][]{BHMIE}. All these methods suffer from discrepancies when compared to DDA or TMM calculations \citep[see Figs.~5 and 10 in][]{Tazaki2018}. In particular, they all fail to reproduce $\kappa_{abs}$ and/or $\kappa_{sca}$ at short wavelengths, implying a poor resemblance of the albedo from the visible to the mid-IR range. To lift these inconsistencies, \citet{Tazaki2018} developed a modified mean field theory (MMF) which accounts for both $\kappa_{abs}$ and $\kappa_{sca}$ at $\lambda \lesssim 20~\mu$m. At longer wavelengths, \citet{Tazaki2018} point out that approximate methods underestimate the far-IR to submm opacity, except the DHS and APMR Mie approaches which slightly overestimate it. The underestimate is simply explained by the fact that in the approximate methods, the emissivity enhancement due to grain-grain contact area in aggregates \citep{Koehler2011} is not taken into account. The overestimate of the DHS and APMR Mie methods is to be nuanced as this mostly depends on the contact surface area chosen in the exact calculations that were used to calibrate the approximate calculations.

\section{Conclusion}
\label{conclusion}

The aim of this paper is to investigate the effects of varying the dust grain structure, the starting point being the perfect compact sphere, on the unpolarised optical properties for grains of a-Sil$_{{\rm Fe,FeS}}$, a-C, and a-C:H. We find that these changes affect the entire wavelength range, from the visible to the far-IR, either by increasing or decreasing the extinction efficiency, spectrally shifting the threshold wavelength above which the absorption efficiency sharply decreases, changing the far-IR dust mass opacity spectral index, all of these effects depend on the material type and the grain structure.

For instance, in the case of isolated compact spherical grains, the threshold wavelength does not only depend on the grain size but also on the insulator/conductor-like behaviour of the material. When porosity is introduced into the grain, whatever the grain composition, for wavelengths shorter than the threshold wavelength $Q_{abs}$ increases while the dust mass opacity increases over the entire wavelength range. For materials having the real part $n$ of their refractive index constant in the far-IR (a-Sil$_{{\rm Fe,FeS}}$ and a-C:H), the increase does not depend on the wavelength and varies linearly with $P$. For material with increasing $n$ in the far-IR (a-C), the dust mass opacity becomes flatter as $P$ increases. The influence of porosity on the albedo remains small for all materials. Similarly, non-sphericity has no influence on the albedo. However, it results in an increase in the dust mass opacity at all wavelengths for the three materials considered, this increase being stronger for conductors than for insulators.

In the case of aggregates made of compact spherical monomers, the threshold wavelength depends on $a_V$, the radius of a compact sphere with same mass, whereas the far-IR dust mass opacity increase is a reflection of the porosity induced by coagulation. For a-Sil$_{{\rm Fe,FeS}}$ and a-C grains, the $\kappa$ increase does not depend on the monomer size $a_0$ but only on the number of monomers $N$ and saturates for $N \geqslant 6-8$, which can be explained by shadowing effects of the aggregate cross-sections. For a-C:H aggregates, the subordination of absorption to scattering implies that the far-IR $\kappa$ increase depends on both $N$ and $a_0$. For all grain compositions, the albedo does not depend on $a_0$ but on $a_V$, whereas on the contrary the asymmetry factor of the phase function $g$ depends on $a_0$, with more forward scattering for smaller monomers. If the aggregates are composed of spheroids instead of spheres, both $Q_{abs}$ and $Q_{sca}$ increases in the visible. The strongest differences between aggregates made of spheres and spheroids are found in the far-IR. For prolate monomers, $\kappa$ increases whereas for oblate monomers, it decreases, which can be explained by different surface shadowing effects. Albedo is not significantly affected by the monomer shape from the visible to the near-IR but the asymmetry factor varies by 5 to 25\% in this same wavelength range. Then, in the case of aggregates made of spheres with different radii, the increases in $Q_{abs}$ and $Q_{sca}$ are stronger than in the case where the monomers have the same radius. The increases are enhanced for more numerous and smaller monomers, which can be seen in terms of increasing the aggregate surface irregularity.

Finally, the silicate mid-IR features at 10 and 18~$\mu$m depend on the grain structure. For isolated grains, starting from compact spherical grains, both features are redshifted and broadened for increasing grain size as long as $a \geqslant 1~\mu$m and for increased porosity irrespective of their size. Isolated oblate and prolate grains also see their features redshifted but the 10~$\mu$m feature is broadened whereas the 18~$\mu$m is narrowed. In the case of aggregates of increasing $a_V$, we find again the limit of $a_V \geqslant 1~\mu$m for significant changes in the features to appear. Then, the 10~$\mu$m feature characteristics depend only on $a_V$ whereas the 18~$\mu$m feature depends also on $a_0$. Monomer shapes do not influence the 10~$\mu$m feature but can change both the peak position and width of the 18~$\mu$m feature. Both the peak position and the width/spectral shape of the 18~$\mu$m feature are more dependent on the grain detailed structure than the 10~$\mu$m feature.

Making use of the results of this study, a new realistic aggregate dust model based on THEMIS \citep{Jones2017}, including ice mantles and material mixing, will be the object of a forthcoming paper in line with the model presented in \citet{Koehler2015}.

To conclude, we would like to stress the importance of taking into account the detailed grain structures for optical property calculations. Even if some promising approximate methods do exist, none of them is able to mimic the entire range of variations at all wavelengths induced by structural variations.

\acknowledgements{We would like to thank L. Verstraete for fruitful discussions about dust optical properties. We also thank B. Draine and V. Ossenkopf for making their sofwares for optical property calculations publicly available. Finally, we thank our referee, M. Mishchenko, for his careful reading and interesting remarks and comments that helped to improve the manuscript. This research was, in part, made possible through the EU FP7 funded project DustPedia (grant No. 606847). It was also supported by the Programme National PCMI of CNRS/INSU with INC/INP co-funded by CEA and CNES.}

\bibliography{biblio}

\begin{thebibliography}{77}
\expandafter\ifx\csname natexlab\endcsname\relax\def\natexlab#1{#1}\fi

\bibitem[{{Anderson}(1958)}]{Anderson1958}
{Anderson}, P.~W. 1958, Physical Review, 109, 1492

\bibitem[{{Bazell} \& {Dwek}(1990)}]{Bazell1990}
{Bazell}, D. \& {Dwek}, E. 1990, \apj, 360, 142

\bibitem[{{Berry} \& {Percival}(1986)}]{Berry1986}
{Berry}, M.~V. \& {Percival}, I.~C. 1986, Optica Acta, 33, 577

\bibitem[{{Bohren} \& {Huffman}(1998)}]{BHMIE}
{Bohren}, C.~F. \& {Huffman}, D.~R. 1998, {Absorption and Scattering of Light
  by Small Particles}, 544

\bibitem[{{Botet} {et~al.}(1997){Botet}, {Rannou}, \& {Cabane}}]{Botet1997}
{Botet}, R., {Rannou}, P., \& {Cabane}, M. 1997, \ao, 36, 8791

\bibitem[{{Bouwman} {et~al.}(2001){Bouwman}, {Meeus}, {de Koter}, {Hony},
  {Dominik}, \& {Waters}}]{Bouwman2001}
{Bouwman}, J., {Meeus}, G., {de Koter}, A., {et~al.} 2001, \aap, 375, 950

\bibitem[{{Doner} \& {Liu}(2017)}]{Doner2017}
{Doner}, N. \& {Liu}, F. 2017, \jqsrt, 187, 10

\bibitem[{{Doner} {et~al.}(2017){Doner}, {Liu}, \& {Yon}}]{Doner2017bis}
{Doner}, N., {Liu}, F., \& {Yon}, J. 2017, Aerosol Science Technology, 51, 532

\bibitem[{{Draine}(1988)}]{Draine1988}
{Draine}, B.~T. 1988, \apj, 333, 848

\bibitem[{{Draine} \& {Flatau}(1994)}]{DDA}
{Draine}, B.~T. \& {Flatau}, P.~J. 1994, Journal of the Optical Society of
  America A, 11, 1491

\bibitem[{{Draine} \& {Flatau}(2013)}]{DDAmanual}
{Draine}, B.~T. \& {Flatau}, P.~J. 2013, ArXiv e-prints

\bibitem[{{Fogel} \& {Leung}(1998)}]{Fogel1998}
{Fogel}, M.~E. \& {Leung}, C.~M. 1998, \apj, 501, 175

\bibitem[{{Foster} \& {Goodman}(2006)}]{Foster2006}
{Foster}, J.~B. \& {Goodman}, A.~A. 2006, \apjl, 636, L105

\bibitem[{{Henning}(2010)}]{Henning2010}
{Henning}, T. 2010, \araa, 48, 21

\bibitem[{{Henning} \& {Stognienko}(1996)}]{Henning1996}
{Henning}, T. \& {Stognienko}, R. 1996, \aap, 311, 291

\bibitem[{{Hollenbach}(1989)}]{Hollenbach1989}
{Hollenbach}, D. 1989, in IAU Symposium, Vol. 135, Interstellar Dust, ed. L.~J.
  {Allamandola} \& A.~G.~G.~M. {Tielens}, 227

\bibitem[{{Jones}(1988)}]{Jones1988}
{Jones}, A.~P. 1988, \mnras, 234, 209

\bibitem[{{Jones}(2011)}]{Jones2011}
{Jones}, A.~P. 2011, \aap, 528, A98

\bibitem[{{Jones}(2012{\natexlab{a}})}]{Jones2012a}
{Jones}, A.~P. 2012{\natexlab{a}}, \aap, 540, A1

\bibitem[{{Jones}(2012{\natexlab{b}})}]{Jones2012b}
{Jones}, A.~P. 2012{\natexlab{b}}, \aap, 540, A2

\bibitem[{{Jones}(2012{\natexlab{c}})}]{Jones2012c}
{Jones}, A.~P. 2012{\natexlab{c}}, \aap, 542, A98

\bibitem[{{Jones} {et~al.}(2013){Jones}, {Fanciullo}, {Koehler}, {Verstraete},
  {Guillet}, {Bocchio}, \& {Ysard}}]{Jones2013}
{Jones}, A.~P., {Fanciullo}, L., {Koehler}, M., {et~al.} 2013, \aap, 558, A62

\bibitem[{{Jones} {et~al.}(2017){Jones}, {K{\"o}hler}, {Ysard}, {Bocchio}, \&
  {Verstraete}}]{Jones2017}
{Jones}, A.~P., {K{\"o}hler}, M., {Ysard}, N., {Bocchio}, M., \& {Verstraete},
  L. 2017, \aap, 602, A46

\bibitem[{{Jones} {et~al.}(2016){Jones}, {K{\"o}hler}, {Ysard}, {Dartois},
  {Godard}, \& {Gavilan}}]{Jones2016}
{Jones}, A.~P., {K{\"o}hler}, M., {Ysard}, N., {et~al.} 2016, \aap, 588, A43

\bibitem[{{Kataoka} {et~al.}(2014){Kataoka}, {Okuzumi}, {Tanaka}, \&
  {Nomura}}]{Kataoka2014}
{Kataoka}, A., {Okuzumi}, S., {Tanaka}, H., \& {Nomura}, H. 2014, \aap, 568,
  A42

\bibitem[{{Kemper} {et~al.}(2004){Kemper}, {Vriend}, \& {Tielens}}]{Kemper2004}
{Kemper}, F., {Vriend}, W.~J., \& {Tielens}, A.~G.~G.~M. 2004, \apj, 609, 826

\bibitem[{{Kemppinen} {et~al.}(2015){Kemppinen}, {Nousiainen}, \&
  {Lindqvist}}]{Kemppinen2015}
{Kemppinen}, O., {Nousiainen}, T., \& {Lindqvist}, H. 2015, \jqsrt, 150, 55

\bibitem[{{Kimura} {et~al.}(2016){Kimura}, {Kolokolova}, \&
  {Lebreton}}]{Kimura2016}
{Kimura}, H., {Kolokolova}, K., \& {Lebreton}, J. 2016, {Light Scattering and
  Thermal Emission by Primitive Dust Particles in Planetary Systems, in Light
  Scattering Reviews, Volume 11 (Springer, ISBN: 978-3-662-49536-0)}

\bibitem[{{Kimura} {et~al.}(2003){Kimura}, {Kolokolova}, \&
  {Mann}}]{Kimura2003}
{Kimura}, H., {Kolokolova}, L., \& {Mann}, I. 2003, \aap, 407, L5

\bibitem[{{Kirchschlager} \& {Wolf}(2013)}]{Kirchschlager2013}
{Kirchschlager}, F. \& {Wolf}, S. 2013, \aap, 552, A54

\bibitem[{{Koehler} {et~al.}(2011){Koehler}, {Guillet}, \&
  {Jones}}]{Koehler2011}
{Koehler}, M., {Guillet}, V., \& {Jones}, A. 2011, \aap, 528, A96

\bibitem[{{Koehler} {et~al.}(2014){Koehler}, {Jones}, \& {Ysard}}]{Koehler2014}
{Koehler}, M., {Jones}, A., \& {Ysard}, N. 2014, \aap, 565, L9

\bibitem[{{Koehler} {et~al.}(2015){Koehler}, {Ysard}, \& {Jones}}]{Koehler2015}
{Koehler}, M., {Ysard}, N., \& {Jones}, A.~P. 2015, \aap, 579, A15

\bibitem[{{K{\"o}hler} {et~al.}(2012){K{\"o}hler}, {Stepnik}, {Jones},
  {Guillet}, {Abergel}, {Ristorcelli}, \& {Bernard}}]{Koehler2012}
{K{\"o}hler}, M., {Stepnik}, B., {Jones}, A.~P., {et~al.} 2012, \aap, 548, A61

\bibitem[{{Kozasa} {et~al.}(1992){Kozasa}, {Blum}, \& {Mukai}}]{Kozasa1992}
{Kozasa}, T., {Blum}, J., \& {Mukai}, T. 1992, \aap, 263, 423

\bibitem[{{Lehtinen} \& {Mattila}(1996)}]{Lehtinen1996}
{Lehtinen}, K. \& {Mattila}, K. 1996, \aap, 309, 570

\bibitem[{{Liu} {et~al.}(2015){Liu}, {Yin}, {Hu}, {Jin}, \&
  {Sorensen}}]{Liu2015}
{Liu}, C., {Yin}, Y., {Hu}, F., {Jin}, H., \& {Sorensen}, C.~M. 2015, Aerosol
  Science Technology, 49, 928

\bibitem[{{Mackowski} \& {Mishchenko}(1996)}]{TMM}
{Mackowski}, D.~W. \& {Mishchenko}, M.~I. 1996, Journal of the Optical Society
  of America A, 13, 2266

\bibitem[{{Martin} {et~al.}(2012){Martin}, {Roy}, {Bontemps},
  {Miville-Desch{\^e}nes}, {Ade}, {Bock}, {Chapin}, {Devlin}, {Dicker},
  {Griffin}, {Gundersen}, {Halpern}, {Hargrave}, {Hughes}, {Klein}, {Marsden},
  {Mauskopf}, {Netterfield}, {Olmi}, {Patanchon}, {Rex}, {Scott}, {Semisch},
  {Truch}, {Tucker}, {Tucker}, {Viero}, \& {Wiebe}}]{Martin2012}
{Martin}, P.~G., {Roy}, A., {Bontemps}, S., {et~al.} 2012, \apj, 751, 28

\bibitem[{{Mattila}(1970)}]{Mattila1970}
{Mattila}, K. 1970, \aap, 9, 53

\bibitem[{{Maxwell Garnett}(1904)}]{MG1904}
{Maxwell Garnett}, J. 1904, Royal Society of London Philosophical Transactions
  Series A, 203

\bibitem[{{Min} {et~al.}(2008){Min}, {Hovenier}, {Waters}, \& {de
  Koter}}]{Min2008}
{Min}, M., {Hovenier}, J.~W., {Waters}, L.~B.~F.~M., \& {de Koter}, A. 2008,
  \aap, 489, 135

\bibitem[{{Min} {et~al.}(2016){Min}, {Rab}, {Woitke}, {Dominik}, \&
  {M{\'e}nard}}]{Min2016}
{Min}, M., {Rab}, C., {Woitke}, P., {Dominik}, C., \& {M{\'e}nard}, F. 2016,
  \aap, 585, A13

\bibitem[{{Min} {et~al.}(2007){Min}, {Waters}, {de Koter}, {Hovenier},
  {Keller}, \& {Markwick-Kemper}}]{Min2007}
{Min}, M., {Waters}, L.~B.~F.~M., {de Koter}, A., {et~al.} 2007, \aap, 462, 667

\bibitem[{{Mishchenko} {et~al.}(2016{\natexlab{a}}){Mishchenko}, {Dlugach}, \&
  {Liu}}]{Mishchenko2016b}
{Mishchenko}, M.~I., {Dlugach}, J.~M., \& {Liu}, L. 2016{\natexlab{a}}, \jqsrt,
  178, 284

\bibitem[{{Mishchenko} {et~al.}(2016{\natexlab{b}}){Mishchenko}, {Dlugach},
  {Yurkin}, {Bi}, {Cairns}, {Liu}, {Panetta}, {Travis}, {Yang}, \&
  {Zakharova}}]{Mishchenko2016a}
{Mishchenko}, M.~I., {Dlugach}, J.~M., {Yurkin}, M.~A., {et~al.}
  2016{\natexlab{b}}, \physrep, 632, 1

\bibitem[{{Mishchenko} {et~al.}(1996){Mishchenko}, {Travis}, \&
  {Mackowski}}]{Mishchenko1996}
{Mishchenko}, M.~I., {Travis}, L.~D., \& {Mackowski}, D.~W. 1996, \jqsrt, 55,
  535

\bibitem[{{Mishchenko} \& {Yurkin}(2017)}]{Mishchenko2017}
{Mishchenko}, M.~I. \& {Yurkin}, M.~A. 2017, Optics Letters, 42, 494

\bibitem[{{Mukai} {et~al.}(1992){Mukai}, {Ishimoto}, {Kozasa}, {Blum}, \&
  {Greenberg}}]{Mukai1992}
{Mukai}, T., {Ishimoto}, H., {Kozasa}, T., {Blum}, J., \& {Greenberg}, J.~M.
  1992, \aap, 262, 315

\bibitem[{{Okamoto} \& {Xu}(1998)}]{Okamoto1998}
{Okamoto}, H. \& {Xu}, Y.-l. 1998, Earth, Planets, and Space, 50, 577

\bibitem[{{Ormel} {et~al.}(2011){Ormel}, {Min}, {Tielens}, {Dominik}, \&
  {Paszun}}]{Ormel2011}
{Ormel}, C.~W., {Min}, M., {Tielens}, A.~G.~G.~M., {Dominik}, C., \& {Paszun},
  D. 2011, \aap, 532, A43

\bibitem[{{Ossenkopf}(1993)}]{Ossenkopf1993}
{Ossenkopf}, V. 1993, \aap, 280, 617

\bibitem[{{Pagani} {et~al.}(2010){Pagani}, {Steinacker}, {Bacmann}, {Stutz}, \&
  {Henning}}]{Pagani2010}
{Pagani}, L., {Steinacker}, J., {Bacmann}, A., {Stutz}, A., \& {Henning}, T.
  2010, Science, 329, 1622

\bibitem[{{Paladini}(2014)}]{Paladini2014}
{Paladini}, R. 2014, in Astrophysics and Space Science Proceedings, Vol.~36,
  The Labyrinth of Star Formation, ed. D.~{Stamatellos}, S.~{Goodwin}, \&
  D.~{Ward-Thompson}, 299

\bibitem[{{Purcell} \& {Pennypacker}(1973)}]{Purcell1973}
{Purcell}, E.~M. \& {Pennypacker}, C.~R. 1973, \apj, 186, 705

\bibitem[{{Rawlings} {et~al.}(2013){Rawlings}, {Juvela}, {Lehtinen}, {Mattila},
  \& {Lemke}}]{Rawlings2013}
{Rawlings}, M.~G., {Juvela}, M., {Lehtinen}, K., {Mattila}, K., \& {Lemke}, D.
  2013, \mnras, 428, 2617

\bibitem[{{Remy} {et~al.}(2017){Remy}, {Grenier}, {Marshall}, \&
  {Casandjian}}]{Remy2017}
{Remy}, Q., {Grenier}, I.~A., {Marshall}, D.~J., \& {Casandjian}, J.~M. 2017,
  \aap, 601, A78

\bibitem[{{Ridderstad} \& {Juvela}(2010)}]{Ridderstad2010}
{Ridderstad}, M. \& {Juvela}, M. 2010, \aap, 520, A18

\bibitem[{{Roy} {et~al.}(2013){Roy}, {Martin}, {Polychroni}, {Bontemps},
  {Abergel}, {Andr{\'e}}, {Arzoumanian}, {Di Francesco}, {Hill}, {Konyves},
  {Nguyen-Luong}, {Pezzuto}, {Schneider}, {Testi}, \& {White}}]{Roy2013}
{Roy}, A., {Martin}, P.~G., {Polychroni}, D., {et~al.} 2013, \apj, 763, 55

\bibitem[{{Shen} {et~al.}(2008){Shen}, {Draine}, \& {Johnson}}]{Shen2008}
{Shen}, Y., {Draine}, B.~T., \& {Johnson}, E.~T. 2008, \apj, 689, 260

\bibitem[{{Stepnik} {et~al.}(2003){Stepnik}, {Abergel}, {Bernard}, {Boulanger},
  {Cambr{\'e}sy}, {Giard}, {Jones}, {Lagache}, {Lamarre}, {Meny}, {Pajot}, {Le
  Peintre}, {Ristorcelli}, {Serra}, \& {Torre}}]{Stepnik2003}
{Stepnik}, B., {Abergel}, A., {Bernard}, J.-P., {et~al.} 2003, \aap, 398, 551

\bibitem[{{Stognienko} {et~al.}(1995){Stognienko}, {Henning}, \&
  {Ossenkopf}}]{Stognienko1995}
{Stognienko}, R., {Henning}, T., \& {Ossenkopf}, V. 1995, \aap, 296, 797

\bibitem[{{Suutarinen} {et~al.}(2013){Suutarinen}, {Haikala}, {Harju},
  {Juvela}, {Andr{\'e}}, {Kirk}, {K{\"o}nyves}, \& {White}}]{Suutarinen2013}
{Suutarinen}, A., {Haikala}, L.~K., {Harju}, J., {et~al.} 2013, \aap, 555, A140

\bibitem[{{Tazaki} \& {Tanaka}(2018)}]{Tazaki2018}
{Tazaki}, R. \& {Tanaka}, H. 2018, ArXiv e-prints

\bibitem[{{Tazaki} {et~al.}(2016){Tazaki}, {Tanaka}, {Okuzumi}, {Kataoka}, \&
  {Nomura}}]{Tazaki2016}
{Tazaki}, R., {Tanaka}, H., {Okuzumi}, S., {Kataoka}, A., \& {Nomura}, H. 2016,
  \apj, 823, 70

\bibitem[{{van Breemen} {et~al.}(2011){van Breemen}, {Min}, {Chiar}, {Waters},
  {Kemper}, {Boogert}, {Cami}, {Decin}, {Knez}, {Sloan}, \&
  {Tielens}}]{Breemen2011}
{van Breemen}, J.~M., {Min}, M., {Chiar}, J.~E., {et~al.} 2011, \aap, 526, A152

\bibitem[{{Vorobiev} {et~al.}(2012){Vorobiev}, {Horley}, {Hern{\'a}ndez-Borja},
  {Esparza-Ponce}, {Ram{\'{\i}}rez-Bon}, {Vorobiev}, {P{\'e}rez}, \&
  {Gonz{\'a}lez-Hern{\'a}ndez}}]{Vorobiev2012}
{Vorobiev}, Y.~V., {Horley}, P.~P., {Hern{\'a}ndez-Borja}, J., {et~al.} 2012,
  Nanoscale Research Letters, 7, 483

\bibitem[{{Voshchinnikov} {et~al.}(2000){Voshchinnikov}, {Il'in}, {Henning},
  {Michel}, \& {Farafonov}}]{Voshchinnikov2000}
{Voshchinnikov}, N.~V., {Il'in}, V.~B., {Henning}, T., {Michel}, B., \&
  {Farafonov}, V.~G. 2000, \jqsrt, 65, 877

\bibitem[{{Wakelam} {et~al.}(2017){Wakelam}, {Bron}, {Cazaux}, {Dulieu}, {Gry},
  {Guillard}, {Habart}, {Hornek{\ae}r}, {Morisset}, {Nyman}, {Pirronello},
  {Price}, {Valdivia}, {Vidali}, \& {Watanabe}}]{Wakelam2017}
{Wakelam}, V., {Bron}, E., {Cazaux}, S., {et~al.} 2017, Molecular Astrophysics,
  9, 1

\bibitem[{{Weingartner} \& {Draine}(2001)}]{Weingartner2001}
{Weingartner}, J.~C. \& {Draine}, B.~T. 2001, \apjs, 134, 263

\bibitem[{{Williams}(2005)}]{Williams2005}
{Williams}, D.~A. 2005, in Journal of Physics Conference Series, Vol.~6,
  Journal of Physics Conference Series, 1--17

\bibitem[{{Wu} {et~al.}(2016){Wu}, {Cheng}, {Zheng}, \& {Chen}}]{Wu2016}
{Wu}, Y., {Cheng}, T., {Zheng}, L., \& {Chen}, H. 2016, \jqsrt, 168, 158

\bibitem[{{Xing} \& {Hanner}(1997)}]{Xing1997}
{Xing}, Z. \& {Hanner}, M.~S. 1997, \aap, 324, 805

\bibitem[{{Yon} {et~al.}(2015){Yon}, {Bescond}, \& {Liu}}]{Yon2015}
{Yon}, J., {Bescond}, A., \& {Liu}, F. 2015, \jqsrt, 162, 197

\bibitem[{{Ysard} {et~al.}(2013){Ysard}, {Abergel}, {Ristorcelli}, {Juvela},
  {Pagani}, {K{\"o}nyves}, {Spencer}, {White}, \& {Zavagno}}]{Ysard2013}
{Ysard}, N., {Abergel}, A., {Ristorcelli}, I., {et~al.} 2013, \aap, 559, A133

\bibitem[{{Ysard} {et~al.}(2016){Ysard}, {K{\"o}hler}, {Jones}, {Dartois},
  {Godard}, \& {Gavilan}}]{Ysard2016}
{Ysard}, N., {K{\"o}hler}, M., {Jones}, A., {et~al.} 2016, \aap, 588, A44

\bibitem[{{Ysard} {et~al.}(2015){Ysard}, {K{\"o}hler}, {Jones},
  {Miville-Desch{\^e}nes}, {Abergel}, \& {Fanciullo}}]{Ysard2015}
{Ysard}, N., {K{\"o}hler}, M., {Jones}, A., {et~al.} 2015, \aap, 577, A110

\end{thebibliography}

\clearpage

\begin{appendix}

\end{appendix}

\end{document}